\documentclass[journal]{IEEEtran}
\ifCLASSINFOpdf
\else
\fi

\usepackage{prettyref}
\usepackage{amsmath}
\usepackage{amsthm}
\usepackage{amssymb}
\usepackage{graphicx}
\usepackage{comment}
\usepackage{subfigure}
\usepackage{pstricks}
\usepackage{pst-node}
\usepackage{pst-grad}
\usepackage{pstricks-add}
\usepackage{cite}

\newtheorem{PROPDEF}{Proposition}
\newtheorem{CORDEF}{Corollary}
\newtheorem{Assumption}{Assumption}
\usepackage{color}

\begin{document}

\title{Fast Desynchronization For Decentralized Multichannel Medium Access Control}

\author{Nikos~Deligiannis,
        Jo\~ao~F.~C.~Mota, George~Smart, and Yiannis~Andreopoulos
\thanks{This work has been presented in part at the 14th International Conference on Information Processing in Sensor Networks (IPSN '15) \cite{deligiannis2015decentralized}.}
\thanks{N. Deligiannis is with the Department
of Electronics and
Informatics, Vrije Universiteit Brussel, Pleinlaan 2, 1050 Brussels, Belgium, and also with iMinds, Ghent 9050, Belgium (email: ndeligia@etro.vub.ac.be).}
\thanks{J. F. C. Mota, G. Smart, and Y. Andreopoulos are with the Electronic and Electrical Engineering Department, University College London, Roberts Building, Torrington Place, London, WC1E 7JE, UK (e-mail:
\{j.mota, george.smart, i.andreopoulos\}@ucl.ac.uk).}
}


\maketitle

\begin{abstract}
Distributed desynchronization algorithms are key to wireless sensor networks as they allow for medium access control in a decentralized manner. In this paper, we view desynchronization primitives as iterative methods that solve  optimization problems.  In particular, by formalizing a well established desynchronization algorithm as a gradient descent method, we establish novel upper bounds on the number of iterations required to reach convergence. Moreover, by using Nesterov's accelerated gradient method, we propose a novel
desynchronization primitive that provides for faster convergence to the steady
state. Importantly, we propose  a novel algorithm that leads to decentralized time-synchronous multichannel TDMA coordination by formulating this  task as an optimization problem.    Our simulations and experiments on a densely-connected  IEEE 802.15.4-based wireless sensor network  demonstrate that our scheme provides for faster   convergence to the  steady state, robustness to hidden nodes, higher network throughput and comparable power dissipation with respect to the recently standardized IEEE 802.15.4e-2012 time-synchronized channel hopping (TSCH) scheme.   
\end{abstract}


\IEEEpeerreviewmaketitle

\section{Introduction}

\IEEEPARstart{I}{n wireless} sensor networks (WSNs), achieving and maintaining (de)synchronization among the nodes   supports various functionalities, including
 data aggregation,  duty cycling, and cooperative
communications.
In particular,  devising protocols that perform desynchronization at the medium access control (MAC) layer is key in achieving fair TDMA scheduling among the nodes in a  channel \cite{degesys2008towards,watteyne2012openwsn,vilajosana2013realistic,tinka2010decentralized,pagliari2011scalable,BuranapanichkitAndreopoulos}.


In order to extend fair TDMA scheduling to  large-scale  networks, protocols that  achieve \textit{(de)synchronization across multiple channels} \cite{vilajosana2013realistic,tinka2010decentralized} are required. Typical approaches are \textit{infrastructure-based} (i.e., \textit{centralized}), as they  use  a coordination channel and/or node and a global clock (e.g., via a GPS system) \cite{tinka2010decentralized}. Channel hopping is been accepted as a good solution for  MAC-layer
coordination for dense WSN topologies. According to channel hopping, nodes hop between the available channels of the physical layer such that they are not constantly using a channel with excessive interference. Forming the state-of-the-art, the \textit{time-synchronized
channel hopping} (TSCH) \cite{tinka2010decentralized} protocol is now
part of the IEEE 802.15.4e-2012 standard \cite{IEEE802.15.4e-2012}. In TSCH, each node reserves timeslots within the predefined slotframe interval and within the 16 channels of IEEE 802.15.4.  However, filling up the available slots follows an advertising
request-and-acknowledgment (RQ/ACK) process on a coordination channel. This channel is prone to interference and  self-inflicted collisions
when nodes advertise slots  aggressively.
Moreover, when  nodes leave
the network, their slots may remain unoccupied for long periods until
another advertisement process reassigns them to other nodes. This  limits the bandwidth usage per channel and does not allow for fast convergence to the steady  state\footnote{Both \textit{high network throughput} and \textit{quick convergence} are important for WSNs that operate with a periodic wake-up cycle (or are event-triggered) and must quickly converge to a steady operational state and transmit high data volumes before being re-suspended.}. 
It is also important to note that TSCH\  requires a coordinator to maintain global time synchronization  \cite{watteyne2012openwsn,tinka2010decentralized}. 

To achieve \textit{infrastructure-less} (i.e., \textit{decentralized}) WSN MAC-layer coordination, distributed (de)synchro-nization algorithms have  attracted a lot of interest  \cite{tinka2010decentralized,simeOne2008distributed,patel2007desync,pagliari2011scalable,motskin2009lightweight,lien2012anchored,leidenfrost2009firefly,klinglmaye2012selforganizing,hong2005scalable,degesys2008towards,choochaisriArtificialForceField,bojic2012self}. These algorithms are inspired by biological agents modeled as pulse-coupled oscillators (PCOs) \cite{mirollo1990synchronization,pagliari2011scalable,klinglmaye2012selforganizing}, namely, as timing mechanisms  following a periodic pulsing (i.e., beacon packet transmission at the MAC) that is updated via the timings of pulses heard from other nodes. 

§Most  work on distributed  (de)synchronization is based on  the PCO dynamics  model introduced by Mirollo and Strogatz  \cite{mirollo1990synchronization}, and derives several algorithms with  properties of practical relevance to WSN deployments, namely: 
\textit{(i)} limited listening \cite{scaglione2010bioinspired,degesys2008towards,wang2012energy}, a property that is imperative for low energy consumption in wireless transceivers;
\textit{(ii)} solutions amenable to multi-hop network topologies and the existence of hidden nodes \cite{motskin2009lightweight,bojic2012self,degesys2008towards}; \textit{(iii)} solutions scalable to large groups of nodes \cite{hong2005scalable,pagliari2011scalable}; and
\textit{(iv)} modifications that lead to fast convergence to steady state \cite{klinglmaye2012selforganizing,leidenfrost2009firefly,lien2012anchored,wang2012optimal}. PCO-based synchronization methods
have also been interpreted as consensus algorithms for multi-agent systems \cite{papachristodoulou2005synchronization,olfati2007consensus,simeone2007distributed}.
The work in \cite{papachristodoulou2005synchronization} studied synchronization of networked
oscillators under heterogeneous time-delays and varying topologies. In \cite{simeone2007distributed},   the synchronization of networked oscillators was
modeled using coupled discrete-time phase locked loops.  

Regarding the study of the convergence speed of desynchronization algorithms, mostly estimates based on simulations or empirical measurements have been derived. In effect, only lower bounds \cite{klinglmaye2012selforganizing,scaglione2010bioinspired}, order-of-convergence estimates \cite{scaglione2010bioinspired,pagliari2011scalable,degesys2008towards} and operational estimates \cite{buranapanichkit2015convergence} have been established.
However, no upper bounds are currently known for the convergence speed of desynchronization algorithms, despite the fact that  such bounds provide for worst-case guarantees of  time and energy consumption to achieve the state of desynchrony. 
Furthermore, despite the plethora of works on PCOs, the problem of extending distributed (de)synchronization algorithms to the multichannel case (which is key in today's wireless networks) has received  limited attention.
A preliminary attempt was done in  \cite{BuranapanichkitAndreopoulos}, where
desynchronization  was independently
applied per channel.  
The limitation of the scheme in \cite{BuranapanichkitAndreopoulos} is that,
since the nodes in different channels are not synchronized, when a node switches channels convergence needs to be established anew.

In this work, we view the problem of desynchronization as an optimization problem. In particular, we show that a minor modification of the well
established \textsc{Desync} algorithm \cite{degesys2008towards,patel2007desync}
is the gradient descent method applied to a specific optimization problem.  Although  desynchronization can also be viewed from a  consensus perspective \cite{olfati2007consensus}, the optimization approach is more powerful as it allows deriving faster algorithms \cite{Erseghe11-FastConsensusByADMM,Mota12-ConsensusOnColoredNetworks-CDC}. Our contributions are as follows:   
\begin{itemize}
\item
We establish novel upper bounds on the convergence
rate of the \textsc{Desync} process. 
Such bounds can yield reliable estimates of worst-case
energy consumption  and time required for convergence, which are important  for systems
that operate under delay and/or  energy constraints.

\item 
We propose a novel desynchronization algorithm based on Nesterov's accelerated  gradient method~\cite{nesterov1983method,Nesterov04-IntroductoryLecturesConvexOptimization}. We show, both theoretically and experimentally, that the proposed   algorithm leads to faster convergence to steady state than the conventional  \textsc{Desync} algorithm \cite{degesys2008towards,patel2007desync}. 

\item 
We propose a novel distributed multichannel method that \textit{jointly} performs   synchronization across channels and desynchronization
within each channel.  Contrary to  \cite{BuranapanichkitAndreopoulos}, the proposed algorithm leads to  time-synchronous multichannel TDMA coordination (where nodes allocated the same timeslot in adjacent channels are synchronized). In this way, nodes can swap channels (thus, avoiding persistent interference in certain channels and achieving higher connectivity) without the network exiting the steady
state. 
\item 
Finally, via simulations and experiments using a real WSN deployment abiding by the IEEE802.15.4 standard, we show that our approach  leads to decentralized time-synchronous multichannel MAC-layer coordination that achieves higher network throughput compared to the state-of-the-art TSCH \cite{tinka2010decentralized} protocol, while incurring comparable   power consumption.
\end{itemize}

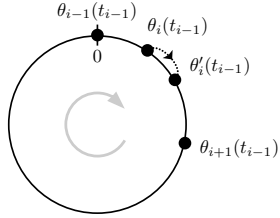
\begin{figure}[t]
         \centering
         \def\scaleFigs{0.7}
         
         \psscalebox{\scaleFigs}{
         \begin{pspicture}(3.8,4.2)

                                                                 \psset{linewidth=0.9pt,labelsep=8pt,arrowsize=7.0pt,arrowinset=0.02}

                 \SpecialCoor

                 \def\radius{1.7}       
                 \def\radiusP{0.12}     
                 \def\radiusA{0.6}      

                 \def\pointSimple{
                         \pscircle*(0,0){\radiusP}
                 }

                 \def\point#1#2{
                                                                        
       \rput(2,2){\rput[lb](\radius;#1){\rnode{x1}{\pointSimple}}}
                    \nput{#1}{x1}{#2}
                 }

                 \pscircle(2,2){\radius}
                 \rput(2,2){\psline(1.5;90)(1.9;90)}
                                                                        \rput(2,2){\rput(\radius;90){\rnode{zero}{}}}
                 \nput{270}{zero}{$0$}

                                                                 \psarc[linecolor=black!20!white,linewidth=1.5pt]{<-}(2,2){\radiusA}{30}{330}

                 \point{-12}{$\theta_{i+1}(t_{i-1})$}
                 \point{90}{$\theta_{i-1}(t_{i-1})$}

                                                                 \rput(2,2){\rput(\radius;56){\rnode{x3k}{\pointSimple}}}
                 \nput{56}{x3k}{$\theta_i(t_{i-1})$}

                                                                 \rput(2,2){\rput(\radius;30){\rnode{x3n}{\pointSimple}}}
                 \nput{30}{x3n}{$\theta_i^\prime(t_{i-1})$}
                                                                 \psset{nodesep=1.0pt,linestyle=dotted,dotsep=0.7pt,arrowsize=7pt,arrowinset=0.2,arrowlength=0.55,linewidth=1.1pt,arcangleA=80,arcangleB=80,ArrowInside=->,ArrowInsidePos=0.56}

                 \ncarc{-}{x3k}{x3n}
         \end{pspicture}
         
         }
         
\caption{
Phase update of node~$i$ according to the \textsc{Desync} algorithm: node~$i-1$
fires at time~$t_{i-1}$, and node~$i$ updates its phase from~$\theta_i(t_{i-1})$
to~$\theta_i^\prime(t_{i-1})$, towards the average of the phases of nodes~$i-1$
and~$i+1$, its phase neighbors.}
\label{fig:DesyncPhaseUpdate}
\end{figure}

The paper continuous as follows: Section \ref{sec:Background} presents the background on PCO methods, while Section \ref{sec:InterpretationAsGradient} derives our upper bound for the desynchronization process and proposes our novel accelerated desynchronization algorithm. Section \ref{sec:MultichannelCoordination} presents  our novel formulation of multichannel coordination. Simulations and experiments using a WSN deployment are given in Section \ref{sec:experiments}, while Section \ref{sec:conclusion} concludes the paper.   

\section{Background on Pulse-Coupled Oscillators }
\label{sec:Background}
Consider a \textit{fully-connected} WSN comprising $n$ nodes, each acting as a pulse-coupled oscillator~\cite{mirollo1990synchronization}. When a node  does not interact with others, it broadcasts a \textit{fire message} or \textit{pulse} periodically. This is modeled by assigning to node $i$ a \textit{phase}~$\theta_i(t)$, whose value at time~$t$ is given by~\cite{degesys2008towards,scaglione2010bioinspired}
\begin{equation}\label{eq:phasenodei}
        \theta_i(t) = \frac{t}{T} + \phi_i \mod{1}\,,
\end{equation}
where~$\phi_i \in [0,1]$ is the \textit{phase offset} of node~$i$ and$\mod{1}$ denotes the modulo operation with respect to unity. 
Fig.~\ref{fig:DesyncPhaseUpdate} illustrates~\eqref{eq:phasenodei} graphically: the phase~$\theta_i(t)$ of node~$i$ can be seen as a bead moving clockwise on a circle, whose origin coincides both with~$0$ and~$1$~\cite{mirollo1990synchronization,refdegdesync,scaglione2010bioinspired,pagliari2011scalable}. If~$\phi_i$ is constant, which happens when the nodes do not interact, node~$i$ broadcasts a fire message every~$T$ time units, when $\theta_i(t) = 1$, and then sets
its phase to zero. When the nodes interact, e.g., by listening to each others' messages, they modify their phases (specifically, their phase offsets), according to an update equation that expresses the PCO dynamics \cite{mirollo1990synchronization}. One of the most prominent PCO  algorithms for desynchronization at the MAC layer of WSNs is the \textsc{Desync} algorithm~\cite{patel2007desync,degesys2008towards}. 
In \textsc{Desync},
the nodes are ordered according to their initial phases: $0 \leq \theta_1(0)
< \theta_2(0) < \cdots < \theta_n(0) < 1$. Assuming perfect
beacon transmission and reception, the order of the firings in \textsc{Desync}
will remain the same \cite{patel2007desync,degesys2008towards}.
The phase $\theta_i$ of each node~$i$ is updated based on the phases $\theta_{i-1}$ and $\theta_{i+1}$ of its \textit{phase neighbors}, nodes $i-1$ and $i+1$, respectively. This is illustrated in Fig.~\ref{fig:DesyncPhaseUpdate}: immediately after node $i-1$ transmits a fire message, node $i$ modifies its phase  according to
\begin{equation}
\theta_{i}^\prime(t_{i-1})=(1-\alpha)\theta_{i}(t_{i-1})+\alpha\frac{\theta_{i-1}(t_{i-1})+\theta_{i+1}(t_{i-1})}{2}\,,
\label{eq:DESYNC_phase_update}
\end{equation}
where~$t_{i-1}$ is the time instant in which node~$i-1$ fires, i.e., $\theta_{i-1}(t_{i-1})=1$,
and~$i=1,2,\dots,n$, with periodic extension at the boundaries. The \textit{jump-phase parameter} $\alpha\in(0,1)$  controls the phase increment~\cite{patel2007desync,degesys2008towards}.

When node~$i$ updates its phase, it has  \textit{stale} knowledge of the phase of node $i+1$, namely, it only knows the previous value of ~$\theta_{i+1}$ and not the current one. This is because node~$i+1$ modified its phase when node~$i$ fired, but the value of the new phase has not been ``announced'' yet~\cite{patel2007desync}.
In \textsc{Desync}, each node: \textit{(i)} updates its phase once in each
\textit{firing round} (we say that a firing round
 is completed when each node in the network
has
fired exactly once); \textit{(ii)}\  does not need to know the total number of nodes,~$n$,
in the network; \textit{(iii)}\ requires \textit{limited listening}, as only
the messages from the two phase neighbors are required. These features make
\textsc{Desync} quite popular~\cite{patel2007desync,degesys2008towards}. 
For a fully-connected network, it has been shown that~\eqref{eq:DESYNC_phase_update} converges  to the \textit{state of desynchrony} at
time~$\overline{t}$,  after which the interval between consecutive firings
is $T/n$ up to a small threshold~$\epsilon$. Under partial connectivity or hidden nodes, convergence is still achieved under a wide variety of topologies, but the node firings may not be equidistant \cite{degesys2008towards}. 
It has been conjectured via simulations \cite{refdegdesync,patel2007desync}
that \textsc{Desync} converges to desynchrony (i.e., perfect TDMA scheduling) in
\begin{equation}
r_{\textsc{Desync}} = O\left(\frac{1}{\alpha}n^{2}\ln\frac{1}{\epsilon}\right)
\label{eq:r_lowbound}
\end{equation}
firing rounds. Recently, under the assumption of uniformly distributed initial firing phases,
an operational estimate for the number of firing rounds for the \textsc{Desync}
algorithm's convergence was derived \cite{buranapanichkit2015convergence}.
However, no upper bounds are known for the desynchronization process.

\begin{figure*}[t]
         \centering
         \def\scaleFigs{0.7}
         \subfigure[]{\label{SubFig:IllustrationStaleA}
         \psscalebox{\scaleFigs}{
         \begin{pspicture}(4.0,4.2)

\psset{linewidth=1.0pt,labelsep=5pt,arrowsize=7.0pt,arrowinset=0.02}

                 \SpecialCoor

                 \def\radius{1.5}       
                 \def\radiusP{0.09}     
                 \def\radiusA{0.6}      

                 \def\pointSimple{
                         \pscircle*(0,0){\radiusP}
                 }

                 \def\point#1#2{
\rput(2,2){\rput(\radius;#1){\rnode{x1}{\pointSimple}}}
                         \nput{#1}{x1}{#2}
                 }

                 \pscircle(2,2){\radius}
                 \rput(2,2){\psline(1.4;90)(1.6;90)}
\rput(2,2){\rput(\radius;90){\rnode{zero}{}}}
                 \nput{270}{zero}{$0$}

\psarc[linecolor=black!20!white,linewidth=1.5pt]{<-}(2,2){\radiusA}{30}{330}

                 \point{100}{\large $\theta_4^{(0)}$}
                 \point{155}{\large $\theta_3^{(0)}$}
                 \point{185}{\large $\theta_2^{(0)}$}
                 \point{235}{\large $\theta_1^{(0)}$}
         \end{pspicture}
         }
         }
         \subfigure[]{\label{SubFig:IllustrationStaleB}
         \psscalebox{\scaleFigs}{
         \begin{pspicture}(3.8,4.2)

\psset{linewidth=1.0pt,labelsep=5pt,arrowsize=7.0pt,arrowinset=0.02}

                 \SpecialCoor

                 \def\radius{1.5}       
                 \def\radiusP{0.09}     
                 \def\radiusA{0.6}      

                 \def\pointSimple{
                         \pscircle*(0,0){\radiusP}
                 }

                 \def\point#1#2{
\rput(2,2){\rput(\radius;#1){\rnode{x1}{\pointSimple}}}
                         \nput{#1}{x1}{#2}
                 }

                 \pscircle(2,2){\radius}
                 \rput(2,2){\psline(1.4;90)(1.6;90)}
\rput(2,2){\rput(\radius;90){\rnode{zero}{}}}
                 \nput{270}{zero}{$0$}

\psarc[linecolor=black!20!white,linewidth=1.5pt]{<-}(2,2){\radiusA}{30}{330}

                 \point{5}{\large $\theta_4^{(0)}$}
                 \point{90}{\large $\theta_2^{(0)}$}
                 \point{140}{\large $\theta_1^{(0)}$}

\rput(2,2){\rput(\radius;60){\rnode{x3k}{\pointSimple}}}
                 \nput{60}{x3k}{\large $\theta_3^{(0)}$}

\rput(2,2){\rput(\radius;30){\rnode{x3n}{\pointSimple}}}
                 \nput{30}{x3n}{\large $\theta_3^{(1)}$}
\psset{nodesep=1.0pt,linestyle=dotted,dotsep=0.7pt,arrowsize=7pt,arrowinset=0.2,arrowlength=0.55,linewidth=1.1pt,arcangleA=80,arcangleB=80,ArrowInside=->,ArrowInsidePos=0.56}

                 \ncarc{-}{x3k}{x3n}
         \end{pspicture}
         }
         }
         \subfigure[]{\label{SubFig:IllustrationStaleC}
         \psscalebox{\scaleFigs}{
         \begin{pspicture}(4.5,4.2)

\psset{linewidth=1.0pt,labelsep=5pt,arrowsize=7.0pt,arrowinset=0.02}

                 \SpecialCoor

                 \def\radius{1.5}       
                 \def\radiusP{0.09}     
                 \def\radiusA{0.6}      

                 \def\pointSimple{
                         \pscircle*(0,0){\radiusP}
                 }

                 \def\point#1#2{
\rput(2,2){\rput(\radius;#1){\rnode{x1}{\pointSimple}}}
                         \nput{#1}{x1}{#2}
                 }

                 \pscircle(2,2){\radius}
                 \rput(2,2){\psline(1.4;90)(1.6;90)}
\rput(2,2){\rput(\radius;90){\rnode{zero}{}}}
                 \nput{270}{zero}{$0$}

\psarc[linecolor=black!20!white,linewidth=1.5pt]{<-}(2,2){\radiusA}{30}{330}


                 \point{275}{\large $\theta_4^{(0)}$}
                 \point{340}{\large $\theta_3^{(1)}$}
                 \point{90}{\large $\theta_1^{(0)}$}

\rput(2,2){\rput(\radius;40){\rnode{x2k}{\pointSimple}}}
                 \nput{40}{x2k}{\large $\theta_2^{(0)}$}

\rput(2,2){\rput(\radius;15){\rnode{x2n}{\pointSimple}}}
                 \nput{15}{x2n}{\large $\theta_2^{(1)}$}
\psset{nodesep=1.0pt,linestyle=dotted,dotsep=0.7pt,arrowsize=7pt,arrowinset=0.2,arrowlength=0.55,linewidth=1.1pt,arcangleA=80,arcangleB=80,ArrowInside=->,ArrowInsidePos=0.56}

                 \ncarc{-}{x2k}{x2n}

         \end{pspicture}
         }
         }
         \subfigure[]{\label{SubFig:IllustrationStaleD}
         \psscalebox{\scaleFigs}{
         \begin{pspicture}(3.7,4.2)

\psset{linewidth=1.0pt,labelsep=5pt,arrowsize=7.0pt,arrowinset=0.02}

                 \SpecialCoor

                 \def\radius{1.5}       
                 \def\radiusP{0.09}     
                 \def\radiusA{0.6}      

                 \def\pointSimple{
                         \pscircle*(0,0){\radiusP}
                 }

                 \def\point#1#2{
\rput(2,2){\rput(\radius;#1){\rnode{x1}{\pointSimple}}}
                         \nput{#1}{x1}{#2}
                 }

                 \pscircle(2,2){\radius}
                 \rput(2,2){\psline(1.4;90)(1.6;90)}
\rput(2,2){\rput(\radius;90){\rnode{zero}{}}}
                 \nput{270}{zero}{$0$}

\psarc[linecolor=black!20!white,linewidth=1.5pt]{<-}(2,2){\radiusA}{30}{330}


                 \point{90}{\large $\theta_4^{(0)}$}
                 \point{155}{\large $\theta_3^{(1)}$}
                 \point{190}{\large $\theta_2^{(1)}$}

\rput(2,2){\rput(\radius;265){\rnode{x1k}{\pointSimple}}}
                 \nput{265}{x1k}{\large $\theta_1^{(0)}$}

\rput(2,2){\rput(\radius;340){\rnode{x1n}{\pointSimple}}}
                 \nput{340}{x1n}{\large $\theta_1^{(1)}$}
\psset{nodesep=1.0pt,linestyle=dotted,dotsep=0.7pt,arrowsize=7pt,arrowinset=0.2,arrowlength=0.55,linewidth=1.1pt,arcangleA=-60,arcangleB=-60,ArrowInside=->,ArrowInsidePos=0.56}

                 \ncarc{-}{x1k}{x1n}
         \end{pspicture}
         }
         }
         \subfigure[]{\label{SubFig:IllustrationStaleE}
         \psscalebox{\scaleFigs}{
         \begin{pspicture}(4.0,4.2)

\psset{linewidth=1.0pt,labelsep=5pt,arrowsize=7.0pt,arrowinset=0.02}

                 \SpecialCoor

                 \def\radius{1.5}       
                 \def\radiusP{0.09}     
                 \def\radiusA{0.6}      

                 \def\pointSimple{
                         \pscircle*(0,0){\radiusP}
                 }
                 \def\pointSimpleG{
                         \pscircle*[linecolor=black!50!white](0,0){\radiusP}
                 }

                 \def\point#1#2{
                                                                                \rput(2,2){\rput(\radius;#1){\rnode{x1}{\pointSimple}}}
                    \nput{#1}{x1}{#2}
                 }
                 
                 \def\pointG#1#2{
                                                                                \rput(2,2){\rput(\radius;#1){\rnode{x1}{\pointSimpleG}}}
                    \nput{#1}{x1}{\color{black!70!white}{#2}}
                 }

                 \pscircle(2,2){\radius}
                 \rput(2,2){\psline(1.4;90)(1.6;90)}
\rput(2,2){\rput(\radius;90){\rnode{zero}{}}}
                 \nput{270}{zero}{$0$}

\psarc[linecolor=black!20!white,linewidth=1.5pt]{<-}(2,2){\radiusA}{30}{330}


                 \point{25}{\large $\theta_4^{(0)}$}
                 \point{90}{\large $\theta_3^{(1)}$}
                 \pointG{146}{\large $\theta_3^{(0)}$}
                 \point{125}{\large $\theta_2^{(1)}$}
                 \point{275}{\large $\theta_1^{(1)}$}

\rput(2,2){\rput(\radius;25){\rnode{x4k}{\pointSimple}}}
                 \nput{25}{x4k}{\large $\theta_4^{(0)}$}

\rput(2,2){\rput(\radius;350){\rnode{x4n}{\pointSimple}}}
                 \nput{350}{x4n}{\large $\theta_4^{(1)}$}
\psset{nodesep=1.0pt,linestyle=dotted,dotsep=0.7pt,arrowsize=7pt,arrowinset=0.2,arrowlength=0.55,linewidth=1.1pt,arcangleA=80,arcangleB=80,ArrowInside=->,ArrowInsidePos=0.56}

                 \ncarc{-}{x4k}{x4n}
         \end{pspicture}
         }
         }
         \caption{
                 Updates during the first iteration of
\textsc{Desync} in a $4$-node network: \text{(a)} initial phases;
no firing has occurred yet; \text{(b)} the first update occurs when
node~$2$ fires and after nodes~$4$ and~$3$ have fired. The firing of
node~$2$ causes node~$3$ to update $\theta_3^{(0)}$ to $\theta_3^{(1)}$. In
the remaining steps, node~$i$ fires and node~$j$  updates its
phase, where~$(i,j)$ is $(1,2)$ in \text{(c)}, $(4,1)$ in \text{(d)},
and $(3,4)$ in \text{(e)}. All phases are updated as a function of the
initial values, i.e., although some of the phases have already changed,
the updates use always $\theta_1^{(0)}$, $\theta_2^{(0)}$, $\theta_3^{(0)}$,
or $\theta_4^{(0)}$, and not the new values.
         }
                     \label{fig:IllustrationStale}
\end{figure*}
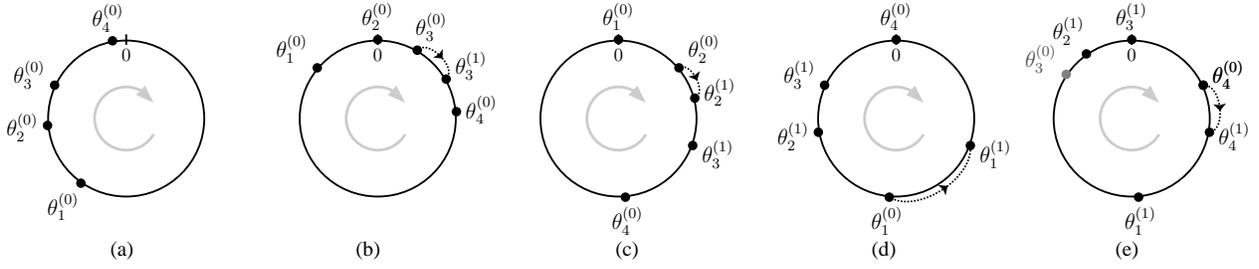

\section{Desync as  a Gradient Method}
\label{sec:InterpretationAsGradient}
We start by showing that, considering a fully-connected network, a minor modification of  \textsc{Desync} \cite{degesys2008towards,patel2007desync}  can be viewed as a gradient descent method solving an optimization problem.
Then, we establish novel convergence properties of the resulting method and derive a new accelerated desynchronization primitive.

\textbf{Staleness of \textsc{Desync}:}
Fig.~\ref{fig:IllustrationStale} shows five consecutive configurations of the phases of the nodes of a network with four nodes. The purpose is to illustrate how the phases of the nodes are updated in the first iteration of \textsc{Desync} \cite{degesys2008towards,patel2007desync} and to highlight our minor modification. For simplicity, we omit the time dependence of the phases, but use a superscript  to indicate how many times they have been updated. In Fig.~\ref{SubFig:IllustrationStaleA}, no firing has yet occurred. The first update occurs when node~$2$ fires, whereby node~$3$ updates its phase from~$\theta_3^{(0)}$ to~$\theta_3^{(1)}$ [see Fig.~\ref{SubFig:IllustrationStaleB}]. According to~\eqref{eq:DESYNC_phase_update}, this update requires knowing~$\theta_2$ (which is equal to~$1$ because node~$2$ is firing) and~$\theta_4$ (which is known because node~$4$ was the first to fire). The second phase update occurs in  Fig.~\ref{SubFig:IllustrationStaleC}: node~$1$ fires, and node~$2$ updates its phase  from~$\theta_2^{(0)}$ to~$\theta_2^{(1)}$. According to~\eqref{eq:DESYNC_phase_update}, this update requires the value of~$\theta_1$ (known because node~$1$ is firing) and~$\theta_3$. The current value of~$\theta_3$ (actually, $\theta_3^{(1)}$) is not known because node~$3$ has not fired since it updated its phase. Therefore, node~$2$ will use~$\theta_3^{(0)}$ rather than~$\theta_3^{(1)}$. This is why we say that \textsc{Desync} is \textit{stale}: each update uses stale versions of the phases. In step~\text{(d)}, node~$1$ updates its phase and also uses a stale version of the phase of node~$2$. Finally, in step~\text{(e)}, node~$4$ updates its phase  using a stale version of the phase of node~$1$. We assume, however, that in contrast with the other nodes, this update uses the value~$\theta_3^{(0)}$ (in gray) and not~$\theta_3^{(1)}$. 
\begin{Assumption}\label{Ass:LastNode}
        In \textsc{Desync}, node~$n$ updates its phase at iteration~$k$ using $\theta_{n-1}^{(k-1)}$ in place of~$\theta_{n-1}^{(k)}$.
\end{Assumption}
Via Assumption \ref{Ass:LastNode}, all updates in Fig.~\ref{fig:IllustrationStale} use the initial values~$\theta_1^{(0)}$, $\theta_2^{(0)}$, $\theta_3^{(0)}$, and~$\theta_4^{(0)}$.
In practice, this assumption does not lead to a discernible difference in the performance of \textsc{Desync}. 

\textbf{Vector notation:}
Suppose we are in the $k$-th firing round, i.e., all nodes have updated their phases~$k-1$ times. We have already mentioned how the firing of a node~$i$, say at time~$t_i$, enables other nodes to determine the current value of~$\phi_i^{(k-1)}$ in~\eqref{eq:phasenodei}: $\phi^{(k-1)}_{i} = 1 - t_i/T$. Knowing this, each node can determine the value of~$\theta_i^{(k-1)}(t)$ for any time instant. We will now see how the update rule~\eqref{eq:DESYNC_phase_update} translates into the updates of the phase offsets. 
Replacing~\eqref{eq:phasenodei} into~\eqref{eq:DESYNC_phase_update} at firing round (iteration)~$k$, we obtain
\begin{align*}
          \theta_i^\prime(t_{i-1}) 
        &= 
          \frac{t_{i-1}}{T} + \phi_i^{(k)}
       \\&=
                (1-\alpha)\Big[\frac{t_{i-1}}{T} + \phi_i^{(k-1)}\Big]
        \\&\qquad +
                \frac{\alpha}{2}\Big[\frac{t_{i-1}}{T} + \phi_{i-1}^{(k-1)} + \frac{t_{i-1}}{T} +  \phi_{i+1}^{(k-1)}\Big]
        \\
        &=
                \frac{t_{i-1}}{T} + (1-\alpha)\phi_i^{(k-1)} + \alpha \frac{\phi_{i-1}^{(k-1)} + \phi_{i+1}^{(k-1)}}{2}\,.
\end{align*}
Eliminating the term~$t_{i-1}/T$, we get: 
$
        \phi_i^{(k)} = (1-\alpha)\phi_i^{(k-1)} + \alpha \frac{\phi_{i-1}^{(k-1)} + \phi_{i+1}^{(k-1)}}{2}\,.
$
In a strict sense, this expression is only valid for $i = 2,\ldots,n-1$ as the updates for nodes~$1$ and~$n$ require a correcting term to compensate the fact that each~$\theta$ wraps around~$1$. Therefore, the updates for all nodes are
\begin{align}
                \phi_1^{(k)} &= (1-\alpha)\phi_1^{(k-1)} + \frac{\alpha}{2}\big(\phi_2^{(k-1)} + \phi_n^{(k-1)} - 1\big)
                \label{Eq:UpdateOffset1}
                \\
                \phi_i^{(k)} &= (1-\alpha)\phi_i^{(k-1)} + \frac{\alpha}{2}\big(\phi_{i-1}^{(k-1)} + \phi_{i+1}^{(k-1)}\big)\,,\, 2\leq i \leq n-1
                \label{Eq:UpdateOffseti}
                \\
                \phi_n^{(k)} &= (1-\alpha)\phi_n^{(k-1)} + \frac{\alpha}{2}\big(\phi_{n-1}^{(k-1)} + \phi_1^{(k-1)} + 1\big)\,.
                \label{Eq:UpdateOffsetn}
\end{align}
Without Assumption~\ref{Ass:LastNode}, $\phi_1^{(k-1)}$ in~\eqref{Eq:UpdateOffsetn} would be replaced with~$\phi_1^{(k)}$. It is, however, this assumption that enables us to write~\eqref{Eq:UpdateOffset1}-\eqref{Eq:UpdateOffsetn} in vector form:
\begin{equation}
\label{eq:DesyncInVector}
      \boldsymbol{\phi}^{(k)} =
      \begin{bmatrix}
          1-\alpha &    \frac{\alpha}{2}     & 0 & \cdots & 0 & \frac{\alpha}{2} \\
          \frac{\alpha}{2} &   1-\alpha      & \frac{\alpha}{2} & \cdots & 0 &  0 \\
          \vdots &  & \ddots & &  \vdots & \vdots \\
          \frac{\alpha}{2} & 0 & 0 & \cdots & \frac{\alpha}{2} &    1-\alpha
      \end{bmatrix}
      \boldsymbol{\phi}^{(k-1)}      
      \\
      -
      \frac{\alpha}{2}\boldsymbol{d}
      \,,
\end{equation}
where $\boldsymbol{\phi}^{(k)} = (\phi_1^{(k)},\phi_2^{(k)},\ldots,\phi_n^{(k)}) \in \mathbb{R}^n$ is a vector containing the phases of all the nodes at iteration~$k$, and $\boldsymbol{d} := (1,0,\ldots,0,-1) \in \mathbb{R}^n$. Equation~\eqref{eq:DesyncInVector} has the format of the updates usually found in the discrete-time consensus literature \cite{DeGroot74-ReachingConsensus,Boyd04-FastLinearIterationsforDistributedAveraging, olfati2007consensus} . In particular, the matrix in~\eqref{eq:DesyncInVector} can be seen as the Perron matrix of a network with a ring topology and the vector $\boldsymbol{d}$ can be seen as an  input bias \cite{olfati2007consensus}. 
This observation can  be used to provide upper bounds on the convergence rate of~\eqref{eq:DesyncInVector}. However, one can view~\eqref{eq:DesyncInVector} as an algorithm solving an optimization problem since, besides also providing upper bounds, this interpretation enables the derivation of an accelerated version of desynchronization. This interpretation is formalized next. 


\begin{PROPDEF} Let $\boldsymbol{\phi}^{(k)}=(\phi_1^{(k)},\phi_2^{(k)},\dots,\phi_n^{(k)})$ denote the phases of all nodes at firing round~$k$. If Assumption~\ref{Ass:LastNode} holds, then \textsc{Desync}~\eqref{eq:DESYNC_phase_update} and~\eqref{eq:DesyncInVector}
is the steepest descent method applied to
\begin{equation}\label{eq:DesyncInterpretation}
     \underset{\boldsymbol{\phi}}{\text{\emph{minimize}}} \,\,\,\, g(\boldsymbol{\phi}) :=
\frac{1}{2}\big\|\boldsymbol{D}\boldsymbol{\phi}-v\boldsymbol{1}_n + \boldsymbol{e}_n\big\|_{2}^{2}
\end{equation}
where $v = 1/n$, $\boldsymbol{1}_{n}\in\mathbb{R}^n$ is the vector of ones, $\boldsymbol{e}_n = (0,0,\ldots,0,1) \in \mathbb{R}^n$, and 
\begin{equation}\label{Eq:MatrixDInDesync}
\boldsymbol{D}=
\begin{bmatrix}
-1 & 1 & 0 & 0 & \dots & 0 \\
0 & -1 & 1 & 0& \dots & 0 \\
\vdots & \ddots &  &  \ddots & & \vdots \\
0 & \dots & 0 & 0 &  -1 & 1 \\
1 & \dots & 0 & 0 &0 & -1 \\
\end{bmatrix}\in \mathbb{R}^{n\times n}\,.
\end{equation}
Specifically, the updates in~\eqref{eq:DesyncInVector} can be written as 
\begin{equation}\label{eq:steepestDescent}
        \boldsymbol{\phi}^{(k)} = \boldsymbol{\phi}^{(k-1)} - \frac{\alpha}{2}\nabla g(\boldsymbol{\phi}^{(k-1)})\,. 
\end{equation} 

\label{prop:DesyncAsGradDescent}
\end{PROPDEF}
\begin{IEEEproof}
        Since $\boldsymbol{D}^{T}\boldsymbol{1}_n=\boldsymbol{0}_{n}$, we have 
\begin{equation}\label{eq:proofDesyncAsGradientStep1}
\nabla g\left( \boldsymbol{\phi}\right)=\boldsymbol{D}^{T}(\boldsymbol{D}\boldsymbol{\phi}-v\boldsymbol{1}_n + \boldsymbol{e}_n)
=
\boldsymbol{D}^{T}\boldsymbol{D}\boldsymbol{\phi} + \boldsymbol{d}\,,
\end{equation} 
where~$\boldsymbol{d} = \boldsymbol{D}^T \boldsymbol{e}_n$ is the vector that appears in~\eqref{eq:DesyncInVector}. Therefore, the steepest descent applied to~\eqref{eq:DesyncInterpretation} yields
\begin{align}
          \boldsymbol{\phi}^{(k)} 
        &= 
                \boldsymbol{\phi}^{(k-1)} - \beta \nabla g(\boldsymbol{\phi}^{(k-1)})
        \nonumber\\
        &=
          \boldsymbol{\phi}^{(k-1)} - \beta \boldsymbol{D}^T \boldsymbol{D} \boldsymbol{\phi}^{(k-1)} - \beta \boldsymbol{d}  
       \nonumber\\
       &=
                (\boldsymbol{I}_n - \beta \boldsymbol{D}^T \boldsymbol{D})\boldsymbol{\phi}^{(k-1)} - \beta \boldsymbol{d}\,,
\label{eq:proofDesyncAsGradientStep1.5}
\end{align}
where $\boldsymbol{I}_n$ is the identity matrix in~$\mathbb{R}^n$. Replacing $\beta = \alpha/2$, we obtain
\begin{equation}\label{eq:proofDesyncAsGradientStep2}
        \boldsymbol{\phi}^{(k)} = (\boldsymbol{I}_n - \frac{\alpha}{2} \boldsymbol{D}^T \boldsymbol{D})\boldsymbol{\phi}^{(k-1)} - \frac{\alpha}{2} \boldsymbol{d}\,,
\end{equation} 
The last equation is exactly~\eqref{eq:DesyncInVector}.
\end{IEEEproof}

We set $v=\frac{1}{n}$ in~\eqref{eq:DesyncInterpretation} to emphasize that the goal of \textsc{Desync} is to disperse the $n$ phases throughout $[0,1]$. However, any other value for~$v$ would lead to the same update rule, since the gradient of the objective function does not depend on~$v$; see~\eqref{eq:proofDesyncAsGradientStep1} in the proof. This confirms the fact that \textsc{Desync} does not require the knowledge of the number of nodes,~$n$, in the network~\cite{patel2007desync}. Notice also that~$\boldsymbol{D}$ is not full rank; therefore, the objective of~\eqref{eq:DesyncInterpretation} is not strictly convex. Indeed, the nullspace of~$\boldsymbol{D}$ is $\{z\,\boldsymbol{1}_n\,:\, z \in \mathbb{R}\} \cup \{\boldsymbol{0}_n\}$. Consequently, if~$\overline{\boldsymbol{\phi}}$ is a solution of~\eqref{eq:DesyncInterpretation}, so is $\overline{\boldsymbol{\phi}} + z\, \boldsymbol{1}_n$ for any~$z \in \mathbb{R}$.
We notice that the interpretation of Proposition 1 is akin to the one that views consensus algorithms as gradient descent methods for minimizing $\sum_{i=1}^{n}(\phi_i-\theta_i)^2$, where $\theta_i$ is the observation of agent $i$ \cite{rabbat2004distributed,Erseghe11-FastConsensusByADMM}.

This interpretation of \textsc{Desync} provides for: \textit{(i)} an alternative way to establish the values of~$\alpha$ for which convergence holds, and \textit{(ii)} an upper bound on the number of the firing rounds until convergence.     
\begin{CORDEF}\label{cor:ConvSteepestDescent}
Every limit point of the sequence produced by the \textsc{Desync} algorithm~\eqref{eq:DesyncInVector} with $\alpha \in (0,1)$ is a stationary point of~\eqref{eq:DesyncInterpretation}. 
\end{CORDEF}   
\begin{IEEEproof}
The proof is given in Appendix \ref{sec:AppendixSimpleGradient}.
\end{IEEEproof}

\begin{CORDEF}\label{cor:RateSteepestDescent}
Let $\boldsymbol{\phi}^{(0)}$ represent the vector of initial phases, and let~$\boldsymbol{\phi}^\star$ be any solution of~\eqref{eq:DesyncInterpretation}. Suppose $\boldsymbol{0}_n\leq\boldsymbol{\phi}^{(0)}\leq \boldsymbol{1}_n$. Then, the number of firing rounds,~$r_{\textsc{D}}$, that \textsc{Desync}~\eqref{Eq:UpdateOffset1}--\eqref{Eq:UpdateOffsetn} requires in order to generate a point~$\overline{\boldsymbol{\phi}}$ that has accuracy $\epsilon:= g\big(\overline{\boldsymbol{\phi}}\big)$ is upper bounded as
\begin{align}
          r_{\textsc{D}} 
        &\leq 
                \frac{\|\boldsymbol{\phi}^{(0)} - \boldsymbol{\phi}^\star\|_2^2}{2\alpha(1-\alpha)}\bigg(\frac{1}{\epsilon} -  \frac{1}{g\big(\boldsymbol{\phi}^{(0)}\big)}\bigg)
        \label{eq:convergenceBoundDesync1}
        \\
        &\leq
                \frac{1}{6n\alpha(1-\alpha)}\left[\frac{7}{2}n^2 + 3n + 4\right]\bigg(\frac{1}{\epsilon} -  \frac{1}{g\big(\boldsymbol{\phi}^{(0)}\big)}\bigg)\,.
        \label{eq:convergenceBoundDesync}
\end{align}
\end{CORDEF}
\begin{IEEEproof}
The proof is given in Appendix \ref{sec:AppendixSimpleGradient}.
\end{IEEEproof}

Corollary 1 confirms Theorem 1 in \cite{patel2007desync} regarding the stability and convergence of \textsc{Desync}, albeit using different tools and without requiring simulations to illustrate the avoidance of limit cycles. Corollary 2 complements the existing order-of-convergence estimate of \eqref{eq:r_lowbound} and the operational estimates derived by Buranapanichkit \textit{et al.} \cite{buranapanichkit2015convergence}  by deriving an upper bound for the firing rounds to achieve convergence.
Such an upper bound allows for reliable estimates of \textit{worst-case} energy consumption  and time, expressed in number of firing rounds or iterations, required to reach convergence.  These estimates are important for systems that operate under delay and/or  energy constraints.
Notice that the bound in~\eqref{eq:convergenceBoundDesync} is a function of known system parameters, namely, the number of nodes $n$, the jump-phase parameter $\alpha$, the tolerance parameter $\epsilon$, and the evaluation of $g(\cdot)$ on the initial phase vector (the latter can be ignored yielding a looser bound). 

\textbf{The \textsc{Fast-Desync} algorithm based on Nesterov:} 
A key advantage of viewing desynchronization as an optimization problem is that we can create new primitives that converge to desynchrony much faster. Particularly, we can use Nesterov's fast gradient algorithm~\cite{Nesterov04-IntroductoryLecturesConvexOptimization,nesterov1983method} (here we use the adaptation in~\cite{Vandenberghe11-Gradient-lecs}):
\begin{subequations}
\label{eq:nesterovDesyncUpdates}
\begin{align}
                \boldsymbol{\phi}^{(k)}
        &=
                \boldsymbol{\mu}^{(k-1)}-\beta\, \nabla g(\boldsymbol{\mu}^{(k-1)})
        \label{eq:nesterovDesyncUpdates:a}
        \\
                \boldsymbol{\mu}^{(k)}
        &=
                \boldsymbol{\phi}^{(k)}+\frac{k-1}{k+2}\big( \boldsymbol{\phi}^{(k)}-\boldsymbol{\phi}^{(k-1)}\big)\,,
        \label{eq:nesterovDesyncUpdates:b}
\end{align}
\end{subequations}
where $\boldsymbol{\mu}^{(k)}\in\mathbb{R}^n$ is an auxiliary vector.
Nesterov's method is applicable under the  same assumptions as the steepest descent, i.e., when~$g$ is continuously differentiable and its gradient is Lipschitz continuous with constant~$L$. However, it requires $0<\beta \leq 1/L$ rather than~$0<\beta < 2/L$. At the expense of small extra memory and computation, Nesterov's method takes $O(1/\sqrt{\epsilon})$ iterations to produce a point~$\overline{\boldsymbol{\phi}}$ that satisfies $g(\overline{\boldsymbol{\phi}}) - g(\boldsymbol{\phi}^\star) \leq \epsilon$, where~$\boldsymbol{\phi}^\star$ minimizes~$g$. Recall that the steepest descent takes~$O(1/\epsilon)$ to produce such a point [cf. \eqref{eq:convergenceBoundDesync}]. We shall show that this improved performance in terms of bounds is also observed experimentally.
Note that~$\boldsymbol{\mu}^{(k)}$,~$\boldsymbol{\phi}^{(k)}$ converge to the same point, i.e.,~$\|\boldsymbol{\phi}^{(k)}-\boldsymbol{\mu}^{(k)}\|\rightarrow{0}$ as $k\rightarrow{\infty}$. More importantly, Nesterov showed in \cite{Nesterov04-IntroductoryLecturesConvexOptimization} that~\eqref{eq:nesterovDesyncUpdates} has optimal convergence rate among first-order methods, i.e., methods that use information about first-order derivatives only, possibly from all past iterations.

 We propose applying Nesterov's algorithm~\eqref{eq:nesterovDesyncUpdates:a}-\eqref{eq:nesterovDesyncUpdates:b} to solve~\eqref{eq:DesyncInterpretation}. This yields a primitive that we call \textsc{Fast-Desync}. Node~$i=1,\ldots,n$ holds two variables~$\phi_i$ and~$\mu_i$, which are updated at iteration~$k$ as
\begin{subequations}
        \begin{align}
                        \phi_i^{(k)} 
                &=
                        (1-\alpha)\mu_i^{(k-1)} + \frac{\alpha}{2}\big(\mu_{i-1}^{(k-1)}+\mu_{i+1}^{(k-1)}-d_i\big)
                \label{eq:nesterovdesyncNode:a}
                \\
                        \mu_i^{(k)}
                &=
                        \phi_i^{(k)} + \frac{k-1}{k+2}\big(\phi_i^{(k)} - \phi_i^{(k-1)}\big)\,,
                \label{eq:nesterovdesyncNode:b}
        \end{align}
\end{subequations}
where~$d_1 = 1$, $d_n = -1$, and~$d_i = 0$ for $i = 2,\ldots,n-1$. Note that~\eqref{eq:nesterovdesyncNode:a} is identical to the \textsc{Desync} updates~\eqref{Eq:UpdateOffset1}--\eqref{Eq:UpdateOffsetn}. The only detriment is that  each node needs an extra memory register to  store~$\phi_i^{(k-1)}$, which is used in~\eqref{eq:nesterovdesyncNode:b}, and perform the extra computations in~\eqref{eq:nesterovdesyncNode:b}. Under this modification, the following holds:
\begin{CORDEF}\label{cor:NesterovDesync}
Let~$\alpha \in (0,1/2]$ and let $\boldsymbol{0}_n \leq \boldsymbol{\phi}^{(0)} = \boldsymbol{\mu}^{(0)} \leq \boldsymbol{1}_n$ represent the vectors of initial phases. Let also~$\boldsymbol{\phi}^\star$ be any solution of~\eqref{eq:DesyncInterpretation}. Then, the number of firing rounds that \textsc{Fast-Desync} \eqref{eq:nesterovdesyncNode:a}-\eqref{eq:nesterovdesyncNode:b} requires to generate a point~$\overline{\boldsymbol{\phi}}$ that has accuracy $\epsilon := g(\overline{\boldsymbol{\phi}})$ is upper bounded as
\begin{align}
          r_{\textsc{FD}} 
        &\leq 
                \frac{2}{\sqrt{\alpha\,\epsilon}}\big\|\boldsymbol{\phi}^{(0)} - \boldsymbol{\phi}^\star\big\|_2
                \label{eq:convergenceBoundNesterovDesyncOptim}
        \\
        &\leq
                2\sqrt{\frac{1}{3n\alpha\epsilon}\left[\frac{7}{2}n^2 + 3n + 4\right]}\,.
                \label{eq:convergenceBoundNesterovDesync}
\end{align}
\end{CORDEF}
\begin{IEEEproof}
The proof is given in Appendix \ref{sec:AppendixSimpleGradient}.
\end{IEEEproof}
Contrasting (\ref{eq:convergenceBoundDesync}) and (\ref{eq:convergenceBoundNesterovDesync}) we notice that  \textsc{Fast-Desync}  allows for significant reduction in the order-of-iterations for convergence compared to   \textsc{Desync}, particularly, $O(\sqrt{n/\epsilon})$ versus $O(n/\epsilon)$, respectively.  

\begin{figure}
\centering
\subfigure[]{
\includegraphics[width=0.48\linewidth]{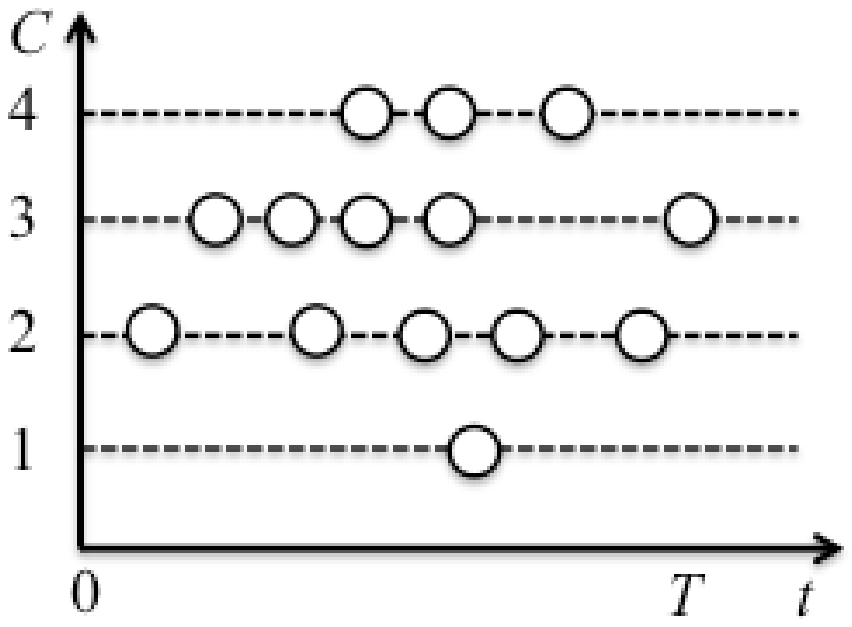}}
\subfigure[]{
\includegraphics[width=0.48\linewidth]{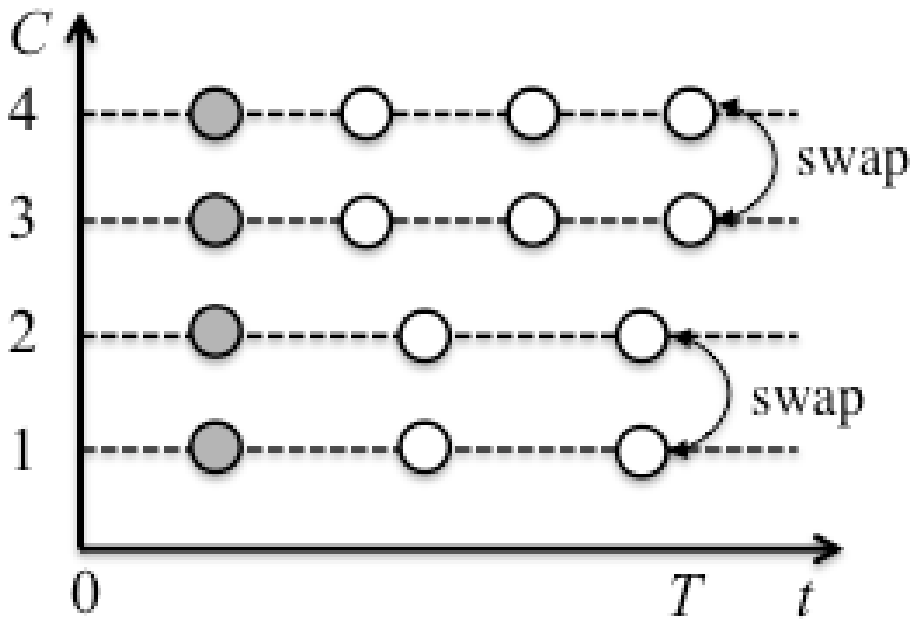}}
\caption{(a) Initial random state  of $n=14$ nodes in $C=4$ channels;
(b) steady state of the proposed protocol with $n_{c}=3$ nodes for channels $c=1$ and $c=2$, and $n_c=4$ nodes in channels $c=3$ and $c=4$. The \textsc{Desync}
nodes (in white) allow for intra-channel
desynchronization, while the \textsc{Sync} nodes (in grey) provide for cross-channel synchronization.  Nodes that belong to balanced channel and that fire synchronously can swap channels.  The horizontal position
of a node indicates the firing moment.   \label{fig:nodesAlignment}}
\end{figure}

\section{Extension To Decentralized Multichannel Coordination}
\label{sec:MultichannelCoordination}
We now describe our  algorithm that jointly  applies synchronization across channels and desynchronization in each channel. We assume that all nodes can receive all fire message broadcasts in their channel. We will show experimentally, however, that our proposal works even for  densely-connected WSNs (when some nodes cannot be reached by others), as \textsc{Desync} still converges in such cases \cite{degesys2008towards}. We first describe our protocol.
 
\subsection{Proposed Decentralized Multichannel MAC-layer Coordination }
\label{sec:protocolIntroduction}
Let a WSN comprise $n$ nodes that are initially  randomly distributed in $C$ channels [see Fig. \ref{fig:nodesAlignment}(a)]---for example, the $C=16$ channels of the IEEE 802.15.4  standard \cite{PerformanceanalysisGTS,wang20136tsch}. The maximum achievable throughput per node is obtained when the nodes are uniformly distributed across the available channels and a perfect TDMA scheduling is reached in each channel.
When the total number of nodes in the network,
$n$, is divisible by $C$ our protocol will lead to  $n_c=\frac{n}{C}$
nodes being present in each
channel, alternatively,  $n_{c}=\left\{ \left\lfloor \frac{n}{C}\right\rfloor
,\left\lceil \frac{n}{C}\right\rceil
\right\}$ nodes will be present in each channel, as shown in Fig. \ref{fig:nodesAlignment}(b). 

Existing  mechanisms, such as the one in   \cite{BuranapanichkitAndreopoulos}, can
 take place during convergence to balance the number of nodes. Specifically, a node lying in channel $c$ may switch to  channel $c+1$ (with cyclic extension at the border), if it detects that less nodes are present there. Detection of the number of nodes in a channel  is possible
by integrating this information in the fire messages transmitted by the nodes. In \cite{BuranapanichkitAndreopoulos}, in order to detect the number of nodes in  channel $c+1$, nodes within channel $c$  proactively switched channels for short time intervals \cite{BuranapanichkitAndreopoulos}.
Here, however, we follow a different  approach, which is akin to the proposed  algorithm. In particular, a single node (which we later call \textsc{Sync}) lying in channel $c$ is elected to listen for fire messages in channel $c+1$. This specific node may jump to the next channel if it detects that less nodes are present there.  When a \textsc{Sync} node jumps from one channel to the next, both
channels are set to elect their \textsc{Sync} nodes anew.
In order to avoid a race condition, where nodes continuously jump channels, the following conditions are defined for channel switching: 
$$
\begin{cases}
n_c-n_{c+1}\geq{1}, \mbox {if}\ c\in[1,C) \\
n_c-n_{c+1}\geq{2}, \mbox {if}\ c=C
\end{cases}
$$
where $n_c$ denotes the number of nodes present in channel $c$, with $n=\sum\limits_{c=1}^{C}n_c$.
The switching
rule and conditions ensure that, after a few firing periods, there will be \(n_{c}\in\{\left\lfloor\frac{n}{C}\right\rfloor,\left\lceil\frac{n}{C}\right\rceil\}\)
nodes in each channel \(c\). 

When the channels have been balanced, the proposed iterative joint synchronization-desynchronization algorithm is applied.  By considering  that each node acts as a pulse-coupled oscillator with a period of  $T$ seconds, our novel algorithm (see Section \ref{sec:proposedJointPrimitive})  leads to \textit{decentralized} multichannel round-robin scheduling. The nodes in each channel are divided in two classes. Specifically, all but one node   in each channel apply desynchronization  so as to achieve TDMA within the channel (these nodes are denoted as \textquotedblleft\textsc{Desync}\textquotedblright). \textsc{Desync} nodes operate only within their channel, firing and listening to  messages from the other nodes in their channel. In addition, one  \textquotedblleft{\textsc{Sync}}\textquotedblright\ node per channel performs cross-channel synchronization to achieve a time-synchronous slot structure  [Fig. \ref{fig:nodesAlignment}(b)].  The \textsc{Sync} node of each channel listens for the \textsc{Sync}
fire message in the next channel\footnote{We consider a cyclic behavior between channels 1 and 16 of IEEE 802.15.4 \cite{wang20136tsch,PerformanceanalysisGTS}. Namely,
the \textsc{Sync} node at channel 16 listens for the fire message from the \textsc{Sync} node in channel 1.}. A node can be designated as the \textsc{Sync} node in a channel based on a pre-established rule, e.g., the node with the smallest node ID, or the node with the highest battery level (all nodes can be made to report their node ID and battery status in their beacon messages).  

We highlight that the existence of a \textsc{Sync} node in each channel calls for an iterative algorithm  performed jointly across the available channels (see Section
\ref{sec:proposedJointPrimitive}). This is fundamentally different from prior schemes, e.g., \cite{BuranapanichkitAndreopoulos}, which applied desynchronization in each channel independently. In contrast, cross-channel synchronization  allows for a \textit{channel swapping} mechanism to be applied  in the converged state.
Specifically, nodes (both of \textsc{Sync} and \textsc{Desync} type) that fire synchronously in adjacent
channels can swap channels and time-slots in pairs
using a simple RQ/ACK scheme\footnote{Swap RQ/ACK packets are transmitted at another channel during a short interval after and before
a node's fire message transmission.} [see Fig. \ref{fig:nodesAlignment}(b)]. Channel swapping allows
for communication between nodes initially present in different channels without leaving the steady network state, thereby achieving increased connectivity. Conversely, in \cite{BuranapanichkitAndreopoulos}, when a node changes channels, convergence to TDMA in the channel needs to be established anew.     

According to our protocol, starting from any random state, the network reaches a steady state, where: \textit{(i)} the same number of nodes is present in adjacent channels, \textit{(ii)} the nodes in each
channel have converged to a TDMA scheduling and \textit{(iii)} the nodes
in  channels with the same number of nodes have a parallel TDMA
scheduling, where nodes allocated with the same time-slot order transmit
synchronously [see Fig.  \ref{fig:nodesAlignment}(b)]. 



\begin{figure}
\begin{centering}
\includegraphics[width=0.7\linewidth]{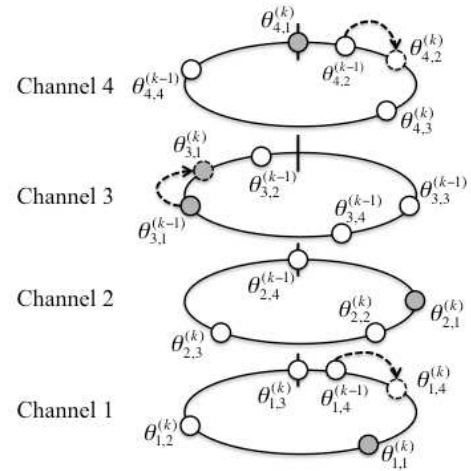}
\par\end{centering}
\caption{Example of the phase updates performed by the proposed multichannel MAC algorithm: In channel 1, the \textsc{Desync} (white) node 4 undergoes a phase update receiving coupling from nodes 1 and 3, present in the same channel.  In channel 2, the \textsc{Sync} (grey) node 2 does \textit{not} receive coupling from the \textsc{Desync} node 4 that fires. In channel 3, the phase of the \textsc{Sync} node 1 is updated due to the firing of the \textsc{Sync} node in channel 4. The firing of the latter node also triggers a phase update of the \textsc{Desync} node 2 in channel 4.}
\label{fig:multichannelCylinder}
\end{figure}

\subsection{Proposed Joint \textsc{Sync-Desync} Algorithm}
\label{sec:proposedJointPrimitive}
We now describe the proposed joint algorithm that allows for synchronization of \textsc{Sync} nodes across channels  and desynchronization of \textsc{Desync} nodes in each channel. Let $\theta_{c,i}$ (resp.\ $\phi_{c,i}$) denote the phase (resp.\ phase offset) of node $i=1,\dots,n_c$ in channel $c=1,\dots,C$. Without loss of generality and  to simplify notation, let the node $i=1$ be the \textsc{Sync} node in each channel\footnote{As explained in Section \ref{sec:protocolIntroduction}, any node in a channel can be the \textsc{Sync} node. This convention is only used to simplify our notation.}.  \textsc{Desync} nodes $i=2,\dots,n_c$ in channel $c$ are coupled with phase neighboring nodes (both \textsc{Desync} and \textsc{Sync}) in the same channel. Namely, any \textsc{Desync} node $i$ in channel $c$ updates its phase offset $\phi_{c,i}$ when node $i-1$ in the same channel transmits a fire message, i.e., when $\theta_{c,i-1}=1$. The \textsc{Sync} node in channel $c$, in turn, receives coupling only from the \textsc{Sync} node in channel $c+1$ (channel 1 for \(c=C\)). Specifically, it updates its phase offset $\phi_{c,1}$ when the \textsc{Sync} node in the next channel fires, that is, when $\theta_{c+1,1}=1$. An illustrative example of the phase updates performed by the proposed algorithm is given in Fig. \ref{fig:multichannelCylinder}. 

\textbf{Problem formulation:} Inspired by the interpretation given in Proposition~\ref{prop:DesyncAsGradDescent}, we address the multichannel coordination problem by solving 
\begin{multline}\label{eq:problemMulichannel}
        \underset{\boldsymbol{\phi_1},\ldots,\boldsymbol{\phi_C}}{\text{minimize}}
        \,\,\,
        h(\boldsymbol{\phi_1},\ldots,\boldsymbol{\phi_C}) := \sum_{c=1}^C         \frac{1}{2}
        \big\|\boldsymbol{D_c} \boldsymbol{\phi_c} - \frac{1}{n_c}\boldsymbol{1}_{n_c} + \boldsymbol{e_{c}}\big\|_2^2
        \\
                +
        \sum_{c=1}^{C}\frac{1}{2}
                \Big(\boldsymbol{w_{c+1}}^T\boldsymbol{\phi}_{c+1} - \boldsymbol{w_{c}}^T\boldsymbol{\phi}_{c}\Big)^2\,,
\end{multline} 
where $\boldsymbol{\phi_c} = (\phi_{c,1},\phi_{c,2},\ldots,\phi_{c,n_c}) \in \mathbb{R}^{n_c}$ is the vector containing the phase offsets of all nodes of channel~$c$, $\boldsymbol{D_c} \in \mathbb{R}^{n_c \times n_c}$ is the  matrix of~\eqref{Eq:MatrixDInDesync} with dimensions $n_c \times n_c$, $\boldsymbol{e_{c}} = (0,0,\ldots,1) \in \mathbb{R}^{n_c}$ and $\boldsymbol{w_c} = (1,0,\ldots,0) \in \mathbb{R}^{n_c}$. 
While the first term of~$h$ enforces desynchronization among the nodes of the same channel [note that each summand has the same format as~$g$ in~\eqref{eq:DesyncInterpretation}], the second term enforces synchronization among the first nodes of each channel. We remark that the second term of \eqref{eq:problemMulichannel} is commonly found in the design of optimization-based consensus algorithms~\cite{Erseghe11-FastConsensusByADMM,rabbat2004distributed,Mota12-ConsensusOnColoredNetworks-CDC}. 

\textbf{Intuition:} We show that the direct application of the gradient descent method to solve~\eqref{eq:problemMulichannel} leads to updates (for the \textsc{Sync} nodes) that cannot be implemented in a practical WSN. However, the proposed  solution will be a modification of those updates.  

Taking into account that~$\boldsymbol{D_c}^T\boldsymbol{1}_{n_c} = \boldsymbol{0}_{n_c}$ for any~$c$, the gradient of~$h$ with respect to~$\boldsymbol{\phi_c}$ is given by
\begin{multline}\label{Eq:GradientMultichannelProblem}
        \nabla_{\boldsymbol{\phi_c}}h(\boldsymbol{\phi_1},\ldots,\boldsymbol{\phi_C})
        =
        \boldsymbol{D_c}^T \boldsymbol{D_c} \boldsymbol{\phi_c} + \boldsymbol{d_c} 
        \\
        + \Big(2\boldsymbol{w_c}^T \boldsymbol{\phi_c} - \boldsymbol{w_{c-1}}^T \boldsymbol{\phi_{c-1}} - \boldsymbol{w_{c+1}}^T \boldsymbol{\phi_{c+1}}\Big)\boldsymbol{w_c}\,,
\end{multline}
where~$\boldsymbol{d_c} := (1,0,\ldots,0,-1) \in \mathbb{R}^{n_c}$. Therefore, the partial derivative of~$h$ with respect to~$\phi_{c,i}$ is
\begin{multline*}
        \frac{\partial}{\partial\, \phi_{c,i}}h(\boldsymbol{\phi_1},\ldots,\boldsymbol{\phi_C})
        \\
        =
        \left\{
                \begin{array}{ll}
                        4\phi_{c,i} - \phi_{c,i-1} - \phi_{c,i+1} - \phi_{c-1,i} - \phi_{c+1,i} &,\,\, i = 1
                        \\
                        2\phi_{c,i} - \phi_{c,i-1} - \phi_{c,i+1} + (\boldsymbol{d_c})_i &,\,\, i \neq 1\,,
                \end{array}
        \right.
\end{multline*} 
where~$(\boldsymbol{d_c})_i$ denotes the $i$-th component of~$\boldsymbol{d_c}$. The gradient descent with stepsize~$\beta$ applied to~\eqref{eq:problemMulichannel} yields for node~$i$ of channel~$c$: $\phi_{c,i}^{(k)} = \phi_{c,i}^{(k-1)} - \beta \frac{\partial}{\partial\, \phi_{c,i}}h(\boldsymbol{\phi_1}^{(k-1)},\ldots,\boldsymbol{\phi_C}^{(k-1)})$. 
Replacing~$\beta$ with~$\alpha/2$, we obtain
\begin{equation}\label{Eq:MultichannelGradientNode1}
        \phi_{c,i}^{(k)} 
        =
        (1-2\alpha)\phi_{c,i}^{(k-1)} + \frac{\alpha}{2}(\phi_{c,i-1}^{(k-1)} + \phi_{c,i+1}^{(k-1)} + \phi_{c-1,i}^{(k-1)} + \phi_{c+1,i}^{(k-1)})\,,
\end{equation} 
for~$i=1$, and
\begin{equation}\label{Eq:MultichannelGradientOtherNodes}
        \phi_{c,i}^{(k)} 
        =
        (1-\alpha)\phi_{c,i}^{(k-1)} + \frac{\alpha}{2}(\phi_{c,i-1}^{(k-1)} + \phi_{c,i+1}^{(k-1)} - (\boldsymbol{d_c})_i)\,,
\end{equation}
for~$i \neq 1$. The update of (\ref{Eq:MultichannelGradientOtherNodes}) is similar  to the \textsc{Desync} algorithm phase update in \eqref{Eq:UpdateOffset1}--\eqref{Eq:UpdateOffsetn}.
However, the derived update for the \textsc{Sync} node, given in~\eqref{Eq:MultichannelGradientNode1}, does not abide by the  coupling rules mentioned in Section~\ref{sec:protocolIntroduction}. Specifically, to implement \eqref{Eq:MultichannelGradientNode1} in a wireless transceiver, each \textsc{Sync} node has to listen for fire messages in its own channel, as well as in the previous and the next channel. This is impractical with the half-duplex transceiver hardware in IEEE 802.15.4-based WSNs.
This issue stems from the symmetry of the matrix~$\boldsymbol{I}_n-\beta \boldsymbol{D}^T \boldsymbol{D}$ [cf. \eqref{Eq:GradientMultichannelProblem} and \eqref{eq:proofDesyncAsGradientStep1.5}]. 
To alleviate this issue, we propose modifying directly the matrix associated with the iterations~\eqref{Eq:MultichannelGradientNode1} and \eqref{Eq:MultichannelGradientOtherNodes}. Our modification is based on the insight that there is one degree of freedom in each channel. Therefore, we can fix the phase of one of the nodes at an arbitrary value. Our approach is to modify~\eqref{Eq:MultichannelGradientNode1}
and \eqref{Eq:MultichannelGradientOtherNodes} to have the first nodes of each channel performing a simple consensus algorithm~\cite{DeGroot74-ReachingConsensus} (while the remaining nodes perform a \textsc{Desync} algorithm). 

\textbf{Multichannel \textsc{Sync-Desync} \textsc{(MuCh-Sync-Desync)}:} For simplicity and without loss of generality, we assume that all channels have the same number of nodes: $n := n_1 = n_2 = \cdots = n_C$. The iteration we propose is
        \begin{multline}\label{Eq:MultiChannelAsymmetricIteration}
                \begin{bmatrix}
                        \boldsymbol{\phi_1}^{(k)} \\
                        \boldsymbol{\phi_2}^{(k)} \\
                        \vdots \\
                        \boldsymbol{\phi_C}^{(k)}
                \end{bmatrix}
                =
                \underbrace{
                \begin{bmatrix}
                        \boldsymbol{Q_1} & \boldsymbol{Q_2} & \boldsymbol{0} & \cdots & \boldsymbol{0}
                        \\
                        \boldsymbol{0} & \boldsymbol{Q_1} & \boldsymbol{Q_2} & \cdots & \boldsymbol{0}
                        \\
                        \vdots & & \ddots &  & \vdots 
                        \\
                        \boldsymbol{0} & \boldsymbol{0} & \cdots & \boldsymbol{Q_1} & \boldsymbol{Q_2}
                        \\
                        \boldsymbol{Q_2} & \boldsymbol{0} & \boldsymbol{0} & \cdots & \boldsymbol{Q_1}
                \end{bmatrix}
                }_{=:\boldsymbol{M}}
                \begin{bmatrix}
                        \boldsymbol{\phi_1}^{(k-1)} \\
                        \boldsymbol{\phi_2}^{(k-1)} \\
                        \vdots \\
                        \boldsymbol{\phi_C}^{(k-1)}
                \end{bmatrix}
                \\
                +
                \underbrace{
                \beta
                \begin{bmatrix}
                        \boldsymbol{e}_n 
                        \\
                        \boldsymbol{e}_n 
                        \\
                        \vdots
                        \\
                        \boldsymbol{e}_n
                \end{bmatrix}
                }_{=:\boldsymbol{b}}
                \,,
        \end{multline}
where~$\boldsymbol{0}$ is the $n\times n$ zero matrix, $\boldsymbol{e}_n:=(0,0\ldots,0,1)\in \mathbb{R}^{n}$, $\boldsymbol{Q_2} := \text{Diag}(\gamma, 0, \ldots, 0)\in \mathbb{R}^{n\times n}$, $0<\gamma<1$, and~$\boldsymbol{Q_1}$ is the $n\times n$ matrix defined as
        $$
                \boldsymbol{Q_1}
                :=
                \begin{bmatrix}
                        1-\gamma & 0 & 0 & 0 & \cdots & 0 & 0 \\                        
                        \beta & 1-2\beta & \beta & 0 & \cdots & 0 & 0 \\
                        0 & \beta & 1-2\beta & \beta & \cdots & 0 & 0 \\
                        \vdots & & & \ddots & & \vdots & \vdots \\
                        \beta & 0 & 0 & 0 & \cdots & \beta & 1-2\beta
                \end{bmatrix}\,.
        $$
        In other words, in each channel~$c$, node~$i\neq 1$  performs the update~\eqref{Eq:MultichannelGradientOtherNodes}, while node~$1$ performs
        \begin{equation}\label{Eq:MultiChannelConsensusNode}
                \phi_{c,1}^{(k)} = (1-\gamma)\phi_{c,1}^{(k-1)} + \gamma\phi_{c+1,1}^{(k-1)}\,.
        \end{equation} 
Recall that the phase update of the \textsc{Sync} node in channel $c$ is performed when the \textsc{Sync} node in channel $c+1$ fires, i.e., when $\theta_{c+1,i}(t_{c+1,i})=1$. Adding $\frac{t_{c+1,i}}{T}$ in both sides of \eqref{Eq:MultiChannelConsensusNode} as well as replacing $\phi_{c+1,i}=1-\frac{t_{c+1,i}}{T}$ and using \eqref{eq:phasenodei} leads to the following phase update for the \textsc{Sync} node in channel $c$:
\begin{equation}
\label{eq:SyncThetaUpdate}
\theta_{c,1}^\prime(t_{c+1,1})= (1-\gamma)\theta_{c,1}(t_{c+1,1})+\gamma\mod{1}.
\end{equation}       
Since $0\leq\theta_{c,1}(t)\leq1$, it is straightforward to show that, for $0<\gamma<1$, \eqref{eq:SyncThetaUpdate} provides for inhibitory coupling\footnote{Similar to other synchronization algorithms \cite{hong2005scalable}, every time the \textsc{Sync} node in channel $c+1$ fires the \textsc{Sync} node in the previous channel will increase its phase towards 1 according to \eqref{eq:SyncThetaUpdate}.} between the \textsc{Sync} nodes in subsequent channels, thereby leading to synchronization of their phases.  In the following proposition we establish that the update  \eqref{Eq:MultiChannelAsymmetricIteration} converges to a solution of the optimization problem~\eqref{eq:problemMulichannel}. In this case, however, we cannot obtain an explicit convergence rate.
Note that the matrix $\boldsymbol{M}$ is not symmetric, which complicates
the convergence analysis. 
Note also that, when the number of nodes per channel varies, the sizes of vectors $\boldsymbol{\phi}$, $\boldsymbol{e}$, and matrices $\boldsymbol{Q_1}$ and $\boldsymbol{Q_2}$ in~\eqref{Eq:MultiChannelAsymmetricIteration} vary per channel $c=1,\dots,C$, but their format is the same. Moreover, the update equations, described in~\eqref{Eq:MultichannelGradientOtherNodes}
and  \eqref{Eq:MultiChannelConsensusNode} remain the same.  
        \begin{PROPDEF}\label{Prop:ConvergenceMultiChannel}
                        Let~$0<\gamma<1$ and~$0<\beta<\frac{1}{2}$. Then, the sequence produced by~\eqref{Eq:MultiChannelAsymmetricIteration} converges to a solution of~\eqref{eq:problemMulichannel}.
        \end{PROPDEF}
        \begin{IEEEproof}
        The proof is given in Appendix~\ref{App:ProofConvAsymmetricMultichannel}.
        \end{IEEEproof}
In  \textsc{MuCh-Sync-Desync}---formed by~\eqref{Eq:MultichannelGradientOtherNodes} and~\eqref{Eq:MultiChannelConsensusNode}---the \textsc{Desync}
and \textsc{Sync} nodes per channel update their phases only once during
a firing round in the channel. Similarly to existing (de)synchronization
algorithms, the role of the parameters~$\alpha$ and~$\gamma$ in the updates of~\eqref{Eq:MultichannelGradientOtherNodes} and~\eqref{Eq:MultiChannelConsensusNode}  is to compensate for missed fire messages and to not allow their propagation throughout all nodes and
channels in the network.        

Since the  update of the \textsc{Desync} nodes in each channel follows the phase update in \eqref{Eq:UpdateOffset1}--\eqref{Eq:UpdateOffsetn}, the corresponding Nesterov modification
can  be applied to speed-up desynchronization in each channel. This approach  leads to the \textsc{Fast-MuCh-Sync-Desync} version of our algorithm, of which the convergence  speed is assessed in the next section.

\section{Experimental Evaluations}
\label{sec:experiments}
\subsection{Simulation Results}

\begin{figure}[t]
\centering
\subfigure[]{
\includegraphics[scale=0.45,trim=0.8cm 0.7cm 1.5cm 0.8cm]{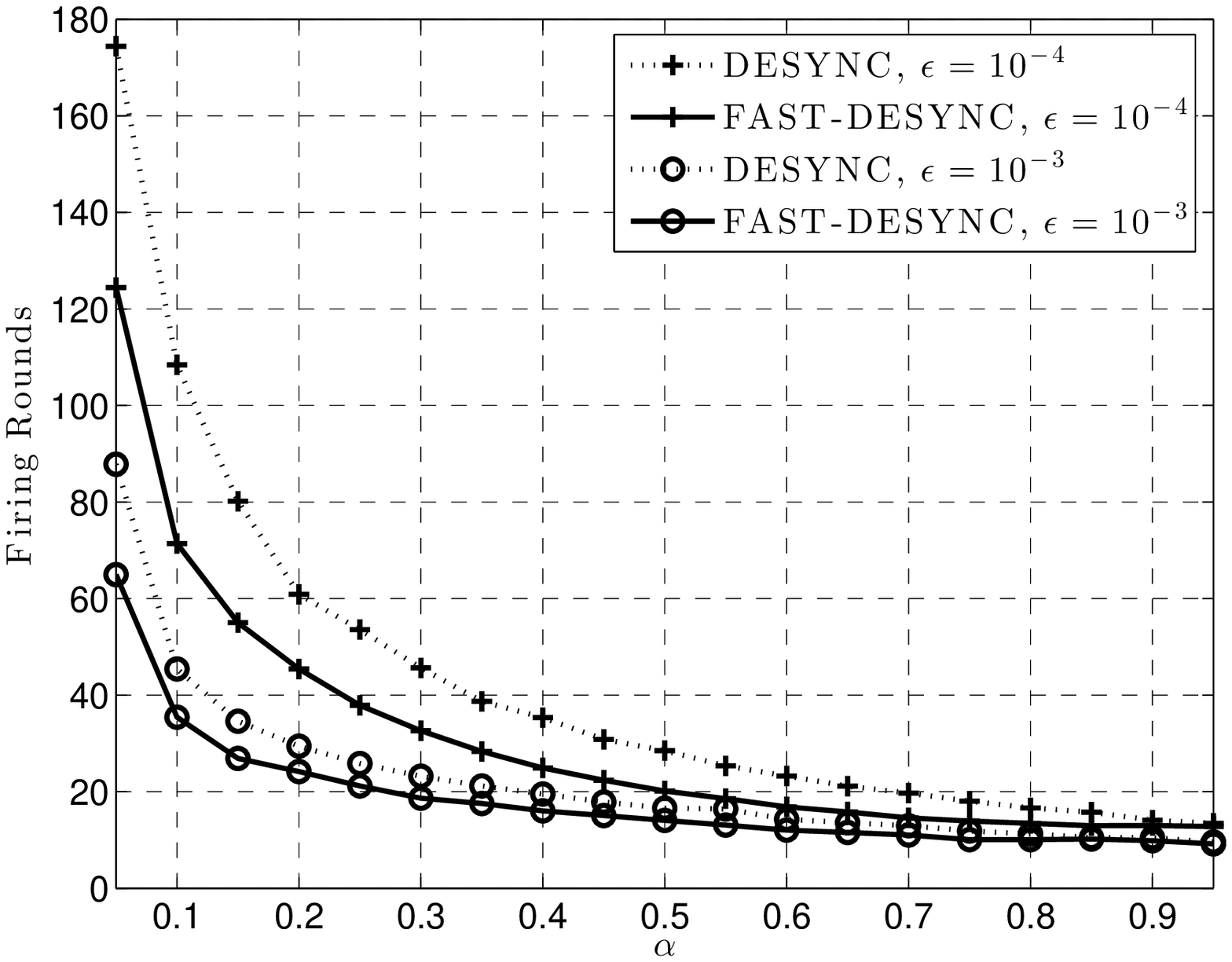}}
\subfigure[]{
\includegraphics[scale=0.45,trim=0.8cm 0.4cm 1.5cm 0.4cm]{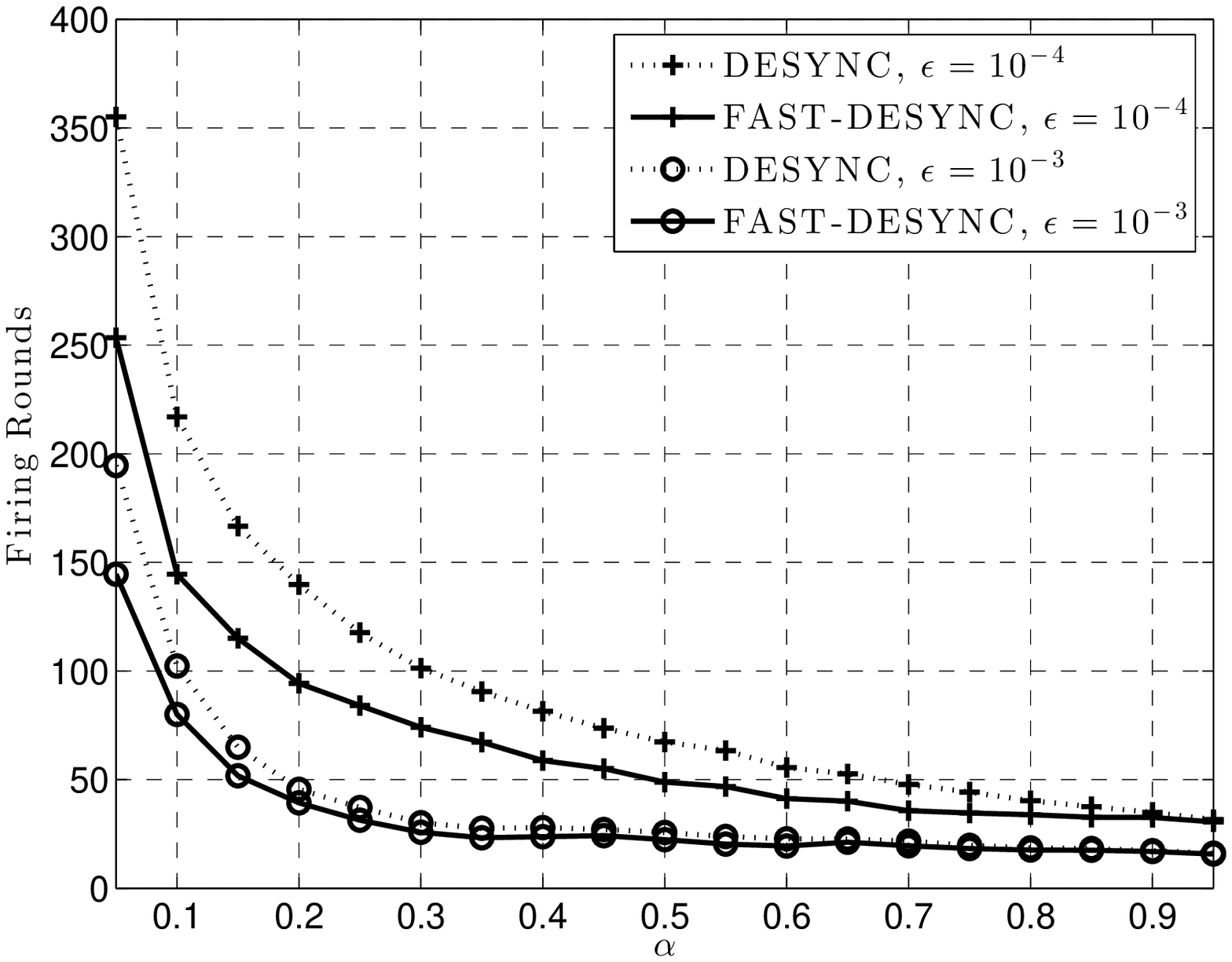}}
\caption{Average number of firing rounds for convergence to TDMA\ scheduling for  \textsc{Desync}  and the proposed \textsc{Fast-Desync} with the Nesterov modification: (a) $n=4$ and (b) $n=8$. \label{fig:DesyncSimulationResults}}
\end{figure}

All simulations  were performed in  MATLAB, by extending the event-driven simulator in \cite{degesys2008towards}. Initially, we examine the performance of \textsc{Desync} versus its fast counterpart based on Nesterov's algorithm. Then, we assess the performance of the proposed \textsc{MuCh-Sync-Desync} algorithm and its fast version. 
We use two  convergence thresholds, i.e., $\epsilon=10^{-3}$ and $\epsilon=10^{-4}$. Convergence is reported at the firing round where the phases $\overline{\boldsymbol{\phi}}$  of the nodes minimize the objective function in \eqref{eq:DesyncInterpretation} with accuracy $g\left(\overline{\boldsymbol{\phi}}\right)\leq\epsilon$. Following existing desynchronization schemes \cite{degesys2008towards,patel2007desync}, our algorithms' updates are performed on the nodes' phases $\theta_i$, as    Assumption 1 does not need  to  be followed in practice. This simplifies the implementation, as we do not need to know the order of  firings.  All simulations were repeated 
400 times and average results are reported.  

\begin{figure}
\begin{centering}
\includegraphics[scale=0.45,trim=0.8cm 0.7cm 1.5cm 0.8cm]{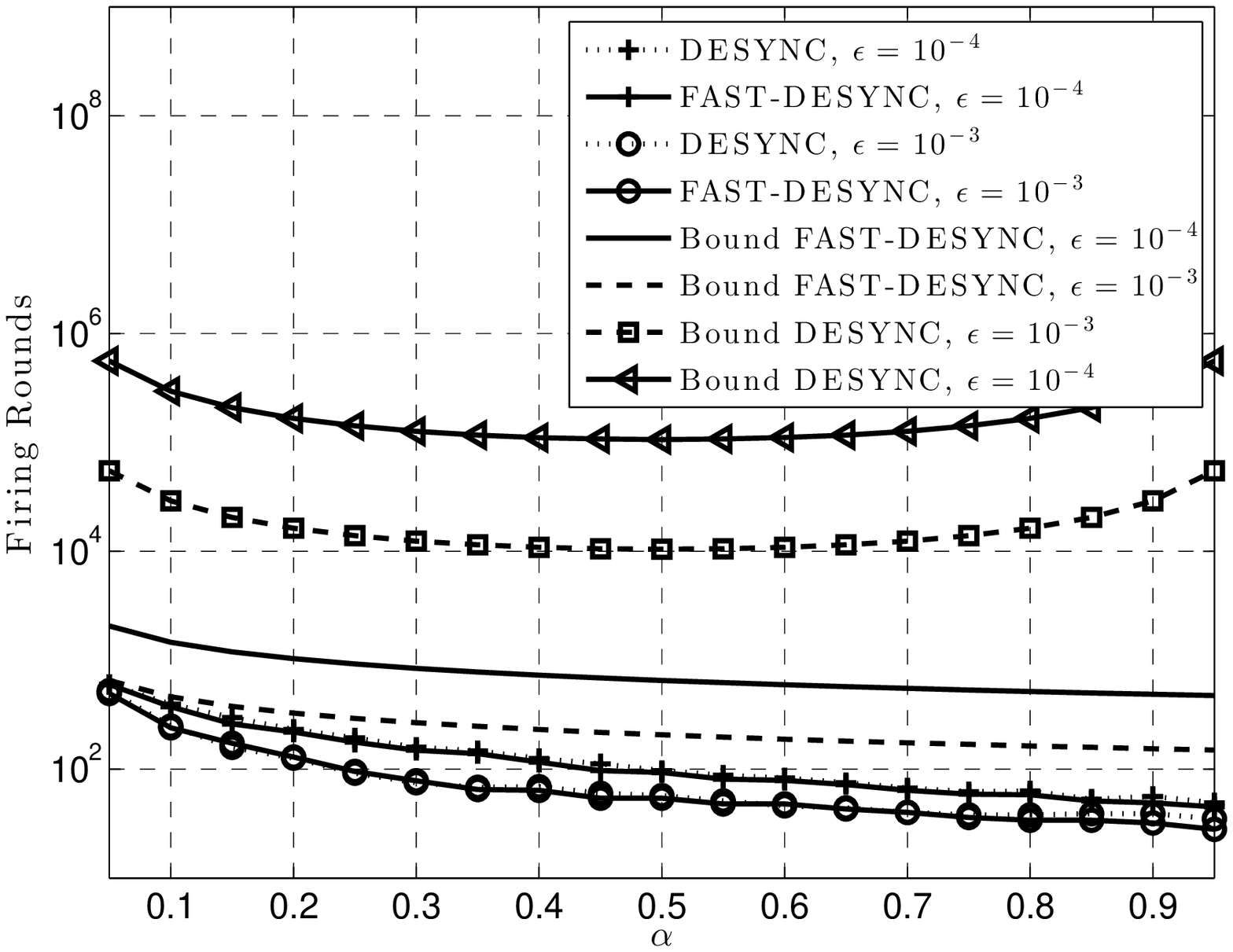}
\par\end{centering}
\caption{Maximum required firing rounds to convergence for \textsc{Desync} and \textsc{Fast-Desync} versus the corresponding upper bounds, $n=8$.}
\label{fig:DesyncSimulationResultsBounds}
\end{figure}


\begin{figure}[t]
\centering
\subfigure[]{
\includegraphics[scale=0.45,trim=0.8cm 0.5cm 1.5cm 0.7cm]{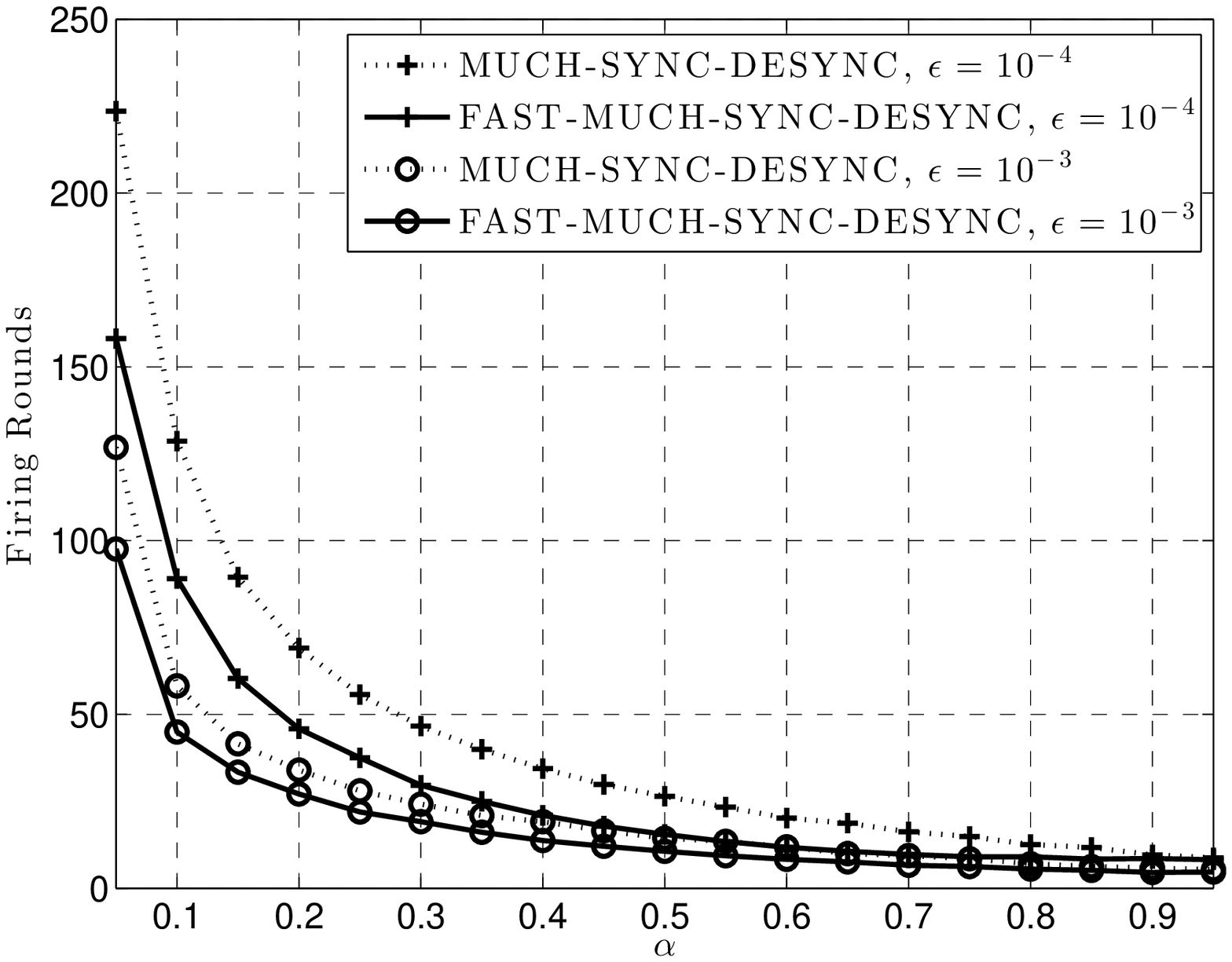}}
\subfigure[]{
\includegraphics[scale=0.45,trim=0.8cm 0.5cm 1.5cm 0.5cm]{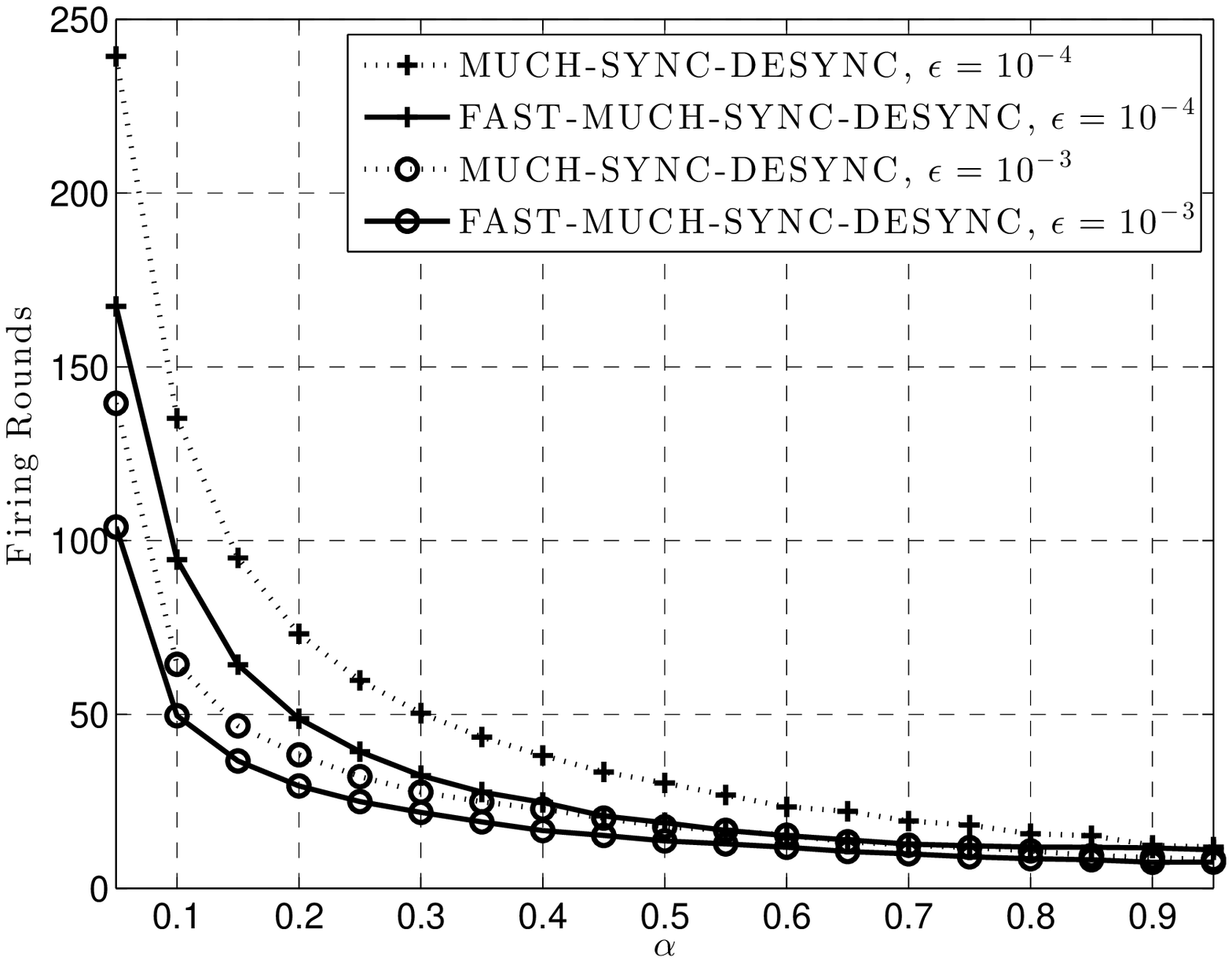}}
\caption{Average number of firing rounds for convergence to decentralized multichannel TDMA scheduling for the proposed \textsc{MuCh-Sync-Desync} algorithm and its fast counterpart; $n_c=4$ nodes per channel are considered with:   (a) $C=6$ and (b) $C=16$ channels. \label{fig:MultiCHSimulationResults}}
\end{figure}

The results of applying desynchronization at a given channel using either  \textsc{Desync} \cite{patel2007desync,degesys2008towards} or the proposed \textsc{Fast-Desync} algorithm are presented in Fig. \ref{fig:DesyncSimulationResults}(a) and (b) for \(n=4\) and \(n=8\) nodes, respectively. Although our analysis proves that \textsc{Fast-Desync}  converges  for $\alpha\in(0,0.5]$,  convergence is actually  achieved for $\alpha\in(0,1)$. In fact, \textsc{Fast-Desync} systematically reduces the required number of iterations to convergence (i.e., irrespective of the value of the parameter $\alpha$), leading to a 2.6\%--28.6\% speed-up with respect to \textsc{Desync}. Furthermore, the convergence speed-up increases when a strict  threshold ($\epsilon=10^{-4}$) is used. The improvement is more significant at low and medium values of $\alpha$, which are typically used in practice  to attenuate the impact of missed fire messages. 

Fig. \ref{fig:DesyncSimulationResultsBounds} depicts the maximum number of
required firing rounds for convergence of \textsc{Desync} and \textsc{Fast-Desync}
versus the   bounds  in Corollaries 2 and 3. The difference between \textsc{Desync}
and \textsc{Fast-Desync} is not visible now  due to the logarithmic scale.
Because of the low $\epsilon$ value in the denominator of \eqref{eq:convergenceBoundDesync}
 the \textsc{Desync} upper bound appears to be loose. ±However, the \textsc{Fast-Desync}
bound in \eqref{eq:convergenceBoundNesterovDesync} offers a tighter characterization
of the simulation-based convergence iterations and follows a trend very similar
to the simulation results.  


We now evaluate the convergence properties of the proposed \textsc{MuCh-Sync-Desync} and its fast version. The results are given in Fig. \ref{fig:MultiCHSimulationResults}(a) and (b) for  $n_{c}=4$ nodes per channel in $C=6$ and $C=16$  channels, respectively. Contrasting these results with the ones  in Fig. \ref{fig:DesyncSimulationResults}, we observe that the proposed multichannel algorithm requires approximately only 10--20\%\ more firing rounds to reach convergence than the single-channel \textsc{Desync} algorithm. It is  also worth noticing that the proposed \textsc{Fast-MuCh-Sync-Desync} version offers a notable convergence speed-up (i.e., 6.01\%--42.54\%) with respect to the simple \textsc{MuCh-Sync-Desync} algorithm,   irrespective of the number of  channels.

\subsection{Experiments with TelosB Motes}

\textbf{Experimental setup:} We implemented the proposed  \textsc{MuCh-Sync-Desync} and its \textsc{Fast} version as applications in the Contiki 2.7 operating system running on TelosB
motes. By utilizing the \texttt{NullMAC} and \texttt{NullRDC} network stack
options in Contiki, we control all node interactions at the MAC layer via
our code. By utilizing the TelosB high-resolution timer (\texttt{rtimer}
library), we can achieve the scheduling of transmission and listening events
with sub-millisecond accuracy, and set  $T=100$ ms. The phase-jump parameters are set as $\alpha=\gamma=0.6$. All nodes first listen constantly until convergence is achieved in their channel, at which point data transmission starts and nodes  switch to sparse listening  to save energy. Due to interference   in the  2.4 GHz band of IEEE 802.15.4 and timing uncertainties in the fire message broadcast and reception, we apply three practical modifications to ensure that, once the network reaches the steady state, it remains there  until the entire network  operation  is suspended, or nodes join or leave the network:  

\begin{enumerate}
\item 
Each node can transmit data in-between its own fire message and the subsequent  fire message from another node, albeit allowing for \textit{guard time} of 6 ms before and after the anticipated beacon broadcast times; this ensures no collisions occur between data and fire message packets.
\item 
In the steady state, each node  turns its transceiver on solely for the 12 ms guard time corresponding to each beacon message.
Moreover, all nodes switch to ``sparse listening'', i.e., they listen for beacons only once every eight periods, unless high interference noise is detected\footnote{In the converged  state, each node determines the interference noise floor in-between transmissions by reading the CC2420 RSSI register. If high interference is detected, the node switches to regular listening. Thus, sparse listening  does not affect the stability of \textsc{MuCh-Sync-Desync}.}.
\item
To remain in sparse listening and avoid interrupting data transmission due to transient interference, all nodes are set to switch to full listening only if  $N_\text{c}=10$ consecutive fire messages are missed. Our choice of \(N_{\text{c}}\) provides  stable operation under interference at the cost of slower reaction
time.
\end{enumerate}
As mentioned in Section \ref{sec:protocolIntroduction}, once all nodes are activated, they are first balanced across the available channels. Note also that, although our time-synchronized slot structure provides  
channel swapping  between synchronous nodes, this is not considered in the experiments.    

We select TSCH  as  benchmark for our comparisons, since it is a state-of-the-art centralized
MAC protocol for densely-connected WSNs \cite{watteyne2012openwsn,vilajosana2013realistic}. Our   implementation follows  the 6tisch simulator and TSCH standard \cite{IEEE802.15.4e-2012,vilajosana2013realistic,wang20136tsch}, namely: channel $11$ of IEEE 802.15.4 was used for advertisements, the RQ/ACK ratio was set to $\frac{1}{9}$, the slotframe comprised $101$ slots of $15$ ms each, and one node was set to broadcast the slotframe beacon for global time synchronization. Finally, the WSN under TSCH is deemed as converged to the steady state when 5\% or less of the timeslots changed within the last 10 slotframes. 

Adhering to scenarios involving  dense network topologies and data-intensive communications (e.g., visual sensor networks \cite{deligiannis2014progressively}), we deployed  $n=64$ nodes in the $C=16$ channels of IEEE 802.15.4. This leads to $n_c=4$ nodes per channel after balancing. The $64$ TelosB motes were placed in four neighboring rooms on the same floor of an office building, with each room containing $16$ nodes.

\textbf{Power dissipation results:}
We assessed the average power dissipation of our scheme against TSCH by placing selected TelosB motes in series with a high-tolerance 1-Ohm resistor and by utilizing a high-frequency oscilloscope to capture the current flow through the resistor in real time. During this experiment, no other devices (or interference signal generators) operating in the 2.4 GHz band were present in the  area. Average results  over $5$ min of operation are reported. The average power dissipation of \textsc{MuCh-Sync-Desync} without transmitting  or receiving data payload was measured to be 1.58 mW. The average power dissipation of a TSCH node under minimal payload (128 bytes per 4 s) was found to be 1.64 mW, which is very close to the value that has been independently reported by Vilajosana \textit{et al.} \cite{vilajosana2013realistic}.
Therefore, under the same setup, our proposal and TSCH were found to incur comparable power dissipation for their operation.   

\textbf{Convergence speed results:}
We  investigate the convergence  time of \textsc{MuCh-Sync-Desync}, \textsc{Fast-MuCh-Sync-Desync} and TSCH under varying interference levels. Rapid convergence to the steady  state is very important when the  WSN is initiated from a suspended state, or when sudden changes happen in the network (e.g.,  nodes join or leave). We carried out 100 independent tests, with each room containing an interference generator for 25 tests.  To generate interference, an RF signal generator was used to create an unmodulated carrier in the center of each WSN channel. The carrier amplitude was adjusted to alter the signal-to-noise-ratio (SNR) at each receiver \cite{boano2011jamlab}. The nodes were set to  maximum transmit power (+0 dBm) in order to operate under the best SNR possible.

\begin{figure}[t]
\begin{centering}
\includegraphics[width=.45\textwidth,trim=0.6cm 0.4cm .9cm 0.9cm]{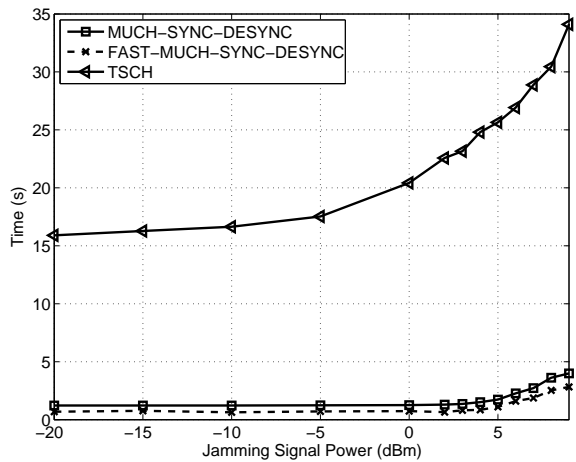}
\par\end{centering}
\caption{Average time required for \textsc{MuCh-Sync-Desync}, its \textsc{Fast}
version, and TSCH to converge under various interference levels.}
\label{fig:TelosBConvergenceResults}
\end{figure}

Fig \ref{fig:TelosBConvergenceResults} shows the time required for \textsc{MuCh-Sync-Desync}, \textsc{Fast-MuCh-Sync-Desync} and TSCH to converge under varying interfering signal power levels. The results corroborate that our proposal reduces the convergence  time by an order of magnitude in comparison to  TSCH and that the Nesterov-based algorithm offers 36.48\%-41.07\% increased convergence speed under a realistic setup. Moreover, the difference in convergence time between the proposed mechanism and TSCH increases with the interference level because TSCH nodes miss most of the  RQ/ACK messages in the advertisement (control) channel. This result demonstrates the key advantages of our decentralized MAC mechanism with respect to TSCH, namely: \textit{(i)} it is fully decentralized and \textit{(ii)} it does not depend on an advertisement and acknowledgement scheme. 

\begin{table}[t]
\caption{Average Convergence Time (in seconds) Under Hidden Nodes. Numbers
in Parenthesis Show the Convergence time of the Fast (Nesterov-based)\ version
of our proposal.} 
\label{tab:convergenceTimeW/WithoutHiddenNodes}
\centering
\begin{tabular}{|c|c|c|}
\hline
 &  \textsc{MuCh-Sync-Desync} & TSCH \\\hline
Without Hidden Nodes & 1.1356 (0.7351) & 15.5845  \\\hline
With Hidden Nodes & 1.8514 (1.2896) & 15.2957  \\\hline
\end{tabular}
\end{table}

\begin{figure}[t]
\begin{centering}
\includegraphics[width=.45\textwidth,trim=0.6cm 0.4cm .9cm 0.9cm]{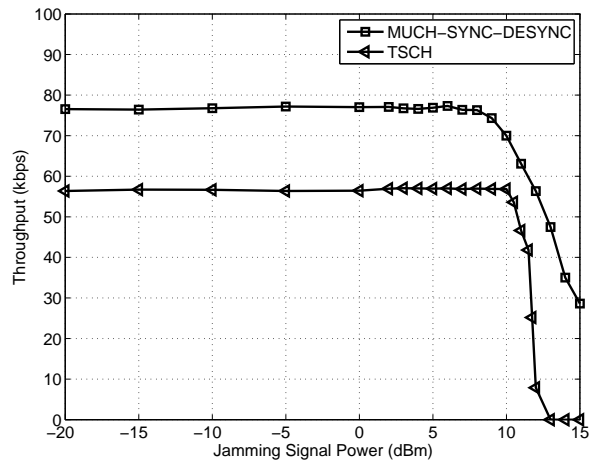}
\par\end{centering}
\caption{Total network throughput  between  \textsc{MuCh-Sync-Desync} and TSCH under varying signal power levels.}
\label{fig:TelosBBandwidthResults}
\end{figure}

\textbf{Results under hidden nodes:}
We now investigate the robustness and convergence speed of our scheme   when some  nodes in the WSN are hidden from other nodes. We measure the time to achieve convergence to steady state when a random subset of $20$ nodes in our WSN setup was programmed to ignore transmissions from $4$ randomly chosen nodes. The results in Table \ref{tab:convergenceTimeW/WithoutHiddenNodes}  show that, irrespective of the presence of hidden nodes, the convergence of  \textsc{MuCh-Sync-Desync} and its \textsc{fast} version is an order-of-magnitude faster than that of TSCH. When hidden nodes are present, the required convergence time of \textsc{MuCh-Sync-Desync} (resp. its \textsc{Fast} version) increases by 63.03\% (resp. 75.43\%), while that of TSCH is actually sightly decreased by 2.13\%. This is to be expected, as TSCH\   nodes simply ignore RQ packets from hidden nodes. Conversely,   due to the \textsc{Desync} (resp. \textsc{Fast-Desync}) process within each channel, applied by \textsc{MuCh-Sync-Desync} (resp. its \textsc{Fast} version), prolonged  beaconing will take place until all hidden nodes are placed amongst non-hidden \textsc{Desync} phase neighbors. This spontaneous robustness of    \textsc{MuCh-Sync-Desync}  (and its \textsc{Fast} version) to hidden nodes is an interesting property that deserves further study\footnote{ For instance, one can try to determine  conditions that guarantee that no configuration of hidden nodes can lead to instability.}.

\textbf{Bandwidth results:}
We measure the total network throughput (i.e., total payload bits transmitted by
all nodes per second) achieved with    \textsc{MuCh-Sync-Desync} and TSCH under various interference levels. Since the measurement is performed after the network is converged, the throughput of   \textsc{MuCh-Sync-Desync} coincides with  its fast version. The results in Fig \ref{fig:TelosBBandwidthResults} show that    \textsc{MuCh-Sync-Desync}   systematically achieves substantially higher  network throughput (more than
40\% increase w.r.t. TSCH), irrespective of the interference level. Both protocols suffer a significant throughput loss of  under high interference  (i.e., above 10 dBm), which is, however, substantially more severe for TSCH. In effect, when interference  is above 12 dBm, the bandwidth obtained with TSCH drops to zero because of the inability to recover lost slots through advertising. Conversely, even under high interference levels,    \textsc{MuCh-Sync-Desync}   recuperates bandwidth utilization due to the elasticity of \textsc{Sync} and \textsc{Desync} mechanisms and the high value used for $N_\text{c}$.


\section{Conclusion}
\label{sec:conclusion}
We have shown that     \textsc{Desync}, which is a well established desynchronization method for MAC layer  coordination in WSNs,   can be viewed as a gradient method for solving an optimization problem. This  interpretation led to a novel, faster desynchronization algorithm (based on Nesterov's modification of the gradient method) and resulted in the derivation of  upper bounds for the convergence of desynchronization. Importantly, casting the problem of time-synchronous desynchronization across channels as a convex optimization problem,   led to the derivation of novel multichannel MAC algorithms. Our proposed \textsc{MuCh-Sync-Desync} algorithm and its fast counterpart were benchmarked against the IEEE 802.15.4e-2012 TSCH\ and were shown to provide for: \textit{(i)} an order-of-magnitude decrease in the convergence time to the network steady state, \textit{(ii)} more than 40\%\ increase in the total network throughput, and \textit{(iii)} significantly-increased robustness to interference and hidden nodes in the network, while requiring comparable power dissipation.   

\appendices

\section{}
\label{sec:AppendixSimpleGradient}
\begin{IEEEproof}[Proof of Corollary \ref{cor:ConvSteepestDescent}]
It is known that every limit point of the steepest descent method with a constant stepsize~$\beta$, i.e.,
$
        \boldsymbol{\phi}^{(k)} = \boldsymbol{\phi}^{(k-1)} - \beta \,\nabla g(\boldsymbol{\phi}^{(k-1)}),
$
is a stationary point of $g(\boldsymbol{\phi})$ whenever~$\nabla g$ is Lipschitz continuous, i.e., there is an $L > 0$ such that
$
\|\nabla g(\boldsymbol{y}) - \nabla g(\boldsymbol{x})\| \leq L\|\boldsymbol{y}
- \boldsymbol{x}\|
$ for all $\boldsymbol{x}$, $\boldsymbol{y} \in \mathbb{R}^n$,
and $\beta \in (0,2/L)$; see~\cite[Prop.1.2.3]{bertsekas1999nonlinear}. In problem (\ref{eq:DesyncInterpretation}), $g$ is twice differentiable,
and $\nabla^2g(\boldsymbol{\phi})=\boldsymbol{D}^T\boldsymbol{D}$, for all
$\boldsymbol{\phi}$. We can then set $L \geq \lambda_{\max}(\boldsymbol{D}^{T}\boldsymbol{D})$,
where $\lambda_{\max}(\cdot)$ is the maximum eigenvalue of a matrix. 
Notice that, for~$\boldsymbol{D}$ in~\eqref{Eq:MatrixDInDesync}, $\boldsymbol{D}^T\boldsymbol{D}$ coincides with the
Laplacian matrix of the ring graph, whose eigenvalues are given by~$2 - 2\cos(2\pi
k/n)$, $k = 1,\ldots,n$ \cite[Lemma 2.4.4]{Spielman-SpectralGrapphTheory-lecs}. We then have
\begin{equation}\label{Eq:AppProofCor1}
\lambda_{\max}(\nabla^2 g(\boldsymbol{\phi}))=\underset{k}{\arg\max} \,\,\, 2 - 2\cos\big(2\pi k/n\big) \leq 4\,.
\end{equation}
Setting~$L=4$, and taking into account that~$\alpha = 2\beta$, we obtain that \textsc{Desync} converges whenever~$\alpha \in (0,1)$. Notice that when~$n$ is even, the maximum is achieved in~\eqref{Eq:AppProofCor1}, i.e., $\lambda_{\max}(\nabla^2 g(\boldsymbol{\phi})) = 4$.
%
\end{IEEEproof}

\begin{IEEEproof}[Proof of Corollary \ref{cor:RateSteepestDescent}]
Let~$g:\mathbb{R}^n\xrightarrow{}\mathbb{R}$ be a convex, continuously differentiable function whose gradient is Lipschitz continuous with constant~$L$. It is known that the sequence generated by the steepest descent method with constant stepsize~$\beta \in (0,2/L)$, i.e.,
$
        \boldsymbol{\phi}^{(k)} = \boldsymbol{\phi}^{(k-1)} - \beta \,\nabla g(\boldsymbol{\phi}^{(k-1)}),
$
satisfies~\cite[Thm.2.1.14]{Nesterov04-IntroductoryLecturesConvexOptimization}
\begin{multline}\label{Eq:AppProofCorollary2Step1}
        g(\boldsymbol{\phi}^{(k)}) - g(\boldsymbol{\phi}^\star)
        \\\leq
        \frac{2(g(\boldsymbol{\phi}^{(0)}) - g(\boldsymbol{\phi}^\star))\|\boldsymbol{\phi}^{(0)} - \boldsymbol{\phi}^\star\|_2^2}{2\|\boldsymbol{\phi}^{(0)} - \boldsymbol{\phi}^\star\|_2^2 + k\,\beta(2-L\beta)(g(\boldsymbol{\phi}^{(0)}) - g(\boldsymbol{\phi}^\star))}\,,
\end{multline} 
where~$\boldsymbol{\phi}^\star$ is any minimizer of~$g$. As shown in the proof of Corollary~\ref{cor:ConvSteepestDescent}, $L=4$ in our case. Furthermore, $g(\boldsymbol{\phi}^\star) = 0$. Taking this into account in~\eqref{Eq:AppProofCorollary2Step1}, using~$\alpha = 2\beta$, and after some manipulations, we get~\eqref{eq:convergenceBoundDesync1}.

To obtain~\eqref{eq:convergenceBoundDesync}, we note that~\eqref{eq:convergenceBoundDesync1} holds for any solution~$\boldsymbol{\phi}^\star$ of~\eqref{eq:DesyncInterpretation}. That is, 
\begin{equation}\label{Eq:AppProofCorollary2Step2}
        r_{\textsc{D}} 
        \leq 
        \bigg[
                \underset{\boldsymbol{\phi}^\star \in S^\star}{\min}\,\, \|\boldsymbol{\phi}^{(0)} - \boldsymbol{\phi}^\star\|_2^2 
        \bigg]\cdot
        \frac{1}{2\alpha(1-\alpha)}\bigg(\frac{1}{\epsilon}-\frac{1}{g\big(\boldsymbol{\phi}^{(0)}\big)}\bigg)\,,
\end{equation} 
where~$S^\star$ is the set of all solutions of~\eqref{eq:DesyncInterpretation}. We have $S^\star = \{\overline{\boldsymbol{\phi}} + z\,\boldsymbol{1}_n\,:\, z \in \mathbb{R}\}$, where~$\overline{\boldsymbol{\phi}}$ is any solution of~\eqref{eq:DesyncInterpretation}. Henceforth, we will take~$\overline{\boldsymbol{\phi}} = (0,1/n,2/n,\ldots,(n-1)/n)$. Then, the minimization problem in~\eqref{Eq:AppProofCorollary2Step2} is equivalent to the minimization of $\|\overline{\boldsymbol{\phi}} +z\,\boldsymbol{1}_n - \boldsymbol{\phi}^{(0)}\|_2^2$ over~$z$, which yields $z^\star = \frac{1}{n}\boldsymbol{1}_n^T(\overline{\boldsymbol{\phi}} - \boldsymbol{\phi}^{(0)})$. Hence,
\begin{align}
                \underset{\boldsymbol{\phi}^\star \in S^\star}{\min}\,\, \|\boldsymbol{\phi}^{(0)} - \boldsymbol{\phi}^\star\|_2^2 
        &=      
                \Big\|\overline{\boldsymbol{\phi}} + \frac{1}{n}\boldsymbol{1}_n^T(\overline{\boldsymbol{\phi}} - \boldsymbol{\phi}^{(0)})\boldsymbol{1}_n - \boldsymbol{\phi}^{(0)}\Big\|_2^2
        \nonumber\\
        &=
                \Big\|\big(\boldsymbol{I}_n + \frac{1}{n}\boldsymbol{1}_n \boldsymbol{1}_n^T\big)\big(\overline{\boldsymbol{\phi}} - \boldsymbol{\phi}^{(0)}\big)\Big\|_2^2\,.
        \label{Eq:AppProofCorollary2Step3}
\end{align} 
To find a worst case scenario, we maximize~\eqref{Eq:AppProofCorollary2Step3} with respect to~$\boldsymbol{\phi}^{(0)}$, subject to the constraints $\boldsymbol{0}_n \leq \boldsymbol{\phi}^{(0)} \leq \boldsymbol{1}_n$. This is a non-convex problem, but the solution can be found in closed-form with the following observation. Since~$\boldsymbol{I}_n + (1/n)\boldsymbol{1}_n \boldsymbol{1}_n^T$ is a circulant matrix and its entries are all positive, maximizing~\eqref{Eq:AppProofCorollary2Step3} subject to $\boldsymbol{0}_n \leq \boldsymbol{\phi}^{(0)} \leq \boldsymbol{1}_n$ is equivalent to
\begin{equation}\label{Eq:AppProofCorollary2Step4}
        \begin{array}{ll}
                \underset{\boldsymbol{\phi}^{(0)}}{\text{maximize}} & \Big\|\overline{\boldsymbol{\phi}} - \boldsymbol{\phi}^{(0)}\Big\|_2^2
                \\
                \text{subject to} & \boldsymbol{0}_n \leq \boldsymbol{\phi}^{(0)} \leq \boldsymbol{1}_n\,.
        \end{array}
\end{equation}
Since~$\overline{\boldsymbol{\phi}} = (0,1/n,2/n,\ldots,(n-1)/n)$, the solution of~\eqref{Eq:AppProofCorollary2Step4} is $\boldmath{\phi}^{(0)} = (1,1,\ldots,1,0,0,\ldots,0)$, where the transition from~$1$ to~$0$ occurs at the first index~$m$ where $m\geq n/2$. Denoting
$$
        B :=
        \begin{array}[t]{cl}
                \underset{\boldsymbol{\phi}^{(0)}}{\max} & \underset{\boldsymbol{\phi}^\star \in S^\star}{\min}\,\, \|\boldsymbol{\phi}^{(0)} - \boldsymbol{\phi}^\star\|_2^2 \\
                \text{s.t.} & \boldsymbol{0}_n \leq \boldsymbol{\phi}^{(0)} \leq \boldsymbol{1}_n
        \end{array}
$$
we have
\begin{align}
          B 
        &\leq
                2\bigg[
          \sum_{i=0}^{m-1}\Big(1 - \frac{i}{n}\Big)^2  + \sum_{i=m}^{n-1}\Big(\frac{i}{n}\Big)^2\bigg]
        \label{Eq:AppProofCorollary2expr1}
        \\
        &=
                \frac{2}{n^2}
                                \bigg[
                                        \sum_{i=0}^{m-1}(n-i)^2 + \sum_{i=m}^{n-1}i^2
                                \bigg]
        \label{Eq:AppProofCorollary2expr2}
        \\
        &=
                \frac{2}{n^2}
                                \bigg[
                                        mn^2 - nm(m-1) + \frac{n(n-1)(2n-1)}{6}
                                \bigg]
        \label{Eq:AppProofCorollary2expr3}
        \\
        &=
                \frac{1}{3n}\left[6nm - 6m^2 + 6m + 2n^2 - 3n + 1\right]
        \label{Eq:AppProofCorollary2expr4}
        \\
        &\leq
                \frac{1}{3n}\left[\frac{7}{2}n^2 + 3n + 4\right].
        \label{Eq:AppProofCorollary2expr5}
\end{align}
The bound in~\eqref{Eq:AppProofCorollary2expr1} is due to replacing     $\boldmath{\phi}^{(0)} = (1,1,\ldots,1,0,0,\ldots,0)$ in~\eqref{Eq:AppProofCorollary2Step3} and using $\big\|(\boldsymbol{I}_n + \frac{1}{n}\boldsymbol{1}_n \boldsymbol{1}_n^T)(\overline{\boldsymbol{\phi}} - \boldsymbol{\phi}^{(0)})\big\|_2^2 \leq \big(\big\|\boldsymbol{I}_n\big\|_2^2 + (1/n^2)\big\|\boldsymbol{1}_n \boldsymbol{1}_n^T\big\|_2^2\big)\big\|\overline{\boldsymbol{\phi}} - \boldsymbol{\phi}^{(0)}\big\|_2^2 = 2\big\|\overline{\boldsymbol{\phi}} - \boldsymbol{\phi}^{(0)}\big\|_2^2$.   From~\eqref{Eq:AppProofCorollary2expr2} to~\eqref{Eq:AppProofCorollary2expr3}, we developed the square in the first summand and used the identities $\sum_{i=1}^{m-1}i = m(m-1)/2$ and $\sum_{i=1}^{n-1}i^2 = n(n-1)(2n-1)/6$. From~\eqref{Eq:AppProofCorollary2expr4} to~\eqref{Eq:AppProofCorollary2expr5}, we used the bound $n/2 \leq m \leq (n+1)/2$. Using~\eqref{Eq:AppProofCorollary2expr5} in~\eqref{Eq:AppProofCorollary2Step2} we get~\eqref{eq:convergenceBoundDesync}.
\end{IEEEproof}

\begin{IEEEproof}[Proof of Corollary \ref{cor:NesterovDesync}]
        Equations~\eqref{eq:nesterovdesyncNode:a}-\eqref{eq:nesterovdesyncNode:b} are applying Nesterov's method~\eqref{eq:nesterovDesyncUpdates:a}-\eqref{eq:nesterovDesyncUpdates:b} to problem~\eqref{eq:DesyncInterpretation} with~$\alpha = 2\beta$. It is known that the number of iterations that~\eqref{eq:nesterovDesyncUpdates:a}-\eqref{eq:nesterovDesyncUpdates:b} requires to generate a point~$\overline{\boldsymbol{\phi}}$ that has accuracy $\epsilon  = g(\overline{\boldsymbol{\phi}})$ is bounded as~\cite{Vandenberghe11-Gradient-lecs}
        \begin{equation}\label{eq:proofNesterovStep1}
                r_{\textsc{FD}} \leq \frac{\sqrt{2/\beta}}{\sqrt{\epsilon - g(\boldsymbol{\phi}^\star)}}\big\|\boldsymbol{\phi}^{(0)} - \boldsymbol{\phi}^\star\big\|_2\,,
        \end{equation} 
        where~$\boldsymbol{\phi}^\star$ minimizes~$g$. This expression is valid for $\beta \in (0,1/L]$, where~$L$ is the Lipschitz constant of~$\nabla g$. We saw in the proof of Corollary~\ref{cor:ConvSteepestDescent} that~$L = 4$ is a valid choice. Since~$g(\boldsymbol{\phi}^\star) = 0$ for any optimal~$\boldsymbol{\phi}^\star$, and using~$\alpha = 2\beta$ in~\eqref{eq:proofNesterovStep1}, we get~\eqref{eq:convergenceBoundNesterovDesyncOptim}. To obtain~\eqref{eq:convergenceBoundNesterovDesync} from~\eqref{eq:convergenceBoundNesterovDesyncOptim}, we  use ~\eqref{Eq:AppProofCorollary2expr5} from the proof of Corollary~\ref{cor:RateSteepestDescent}.
\end{IEEEproof}

\section{}
\label{App:ProofConvAsymmetricMultichannel}

\begin{IEEEproof} [Proof of Proposition \ref{Prop:ConvergenceMultiChannel}]
        If~$\{\boldsymbol{\phi}^{(k)}\}$ converges, its limit will be a fixed point of~\eqref{Eq:MultiChannelAsymmetricIteration}. Before showing that~$\{\boldsymbol{\phi}^{(k)}\}$ converges, we show that any fixed point of~\eqref{Eq:MultiChannelAsymmetricIteration} solves~\eqref{eq:problemMulichannel}. Let~$\boldsymbol{\phi}^\star = (\boldsymbol{\phi}_1^\star, \boldsymbol{\phi}_2^\star, \ldots, \boldsymbol{\phi}_C^\star)$ be a fixed point of~\eqref{Eq:MultiChannelAsymmetricIteration}. For each~$c$, we have
        \begin{align}
                \phi_{c,i}^\star &= (1-\gamma)\phi_{c,i}^\star + \gamma \,\phi_{c+1,i}^\star\,, \,\,\, i=1
                \label{Eq:ProofMultiChannelAsymmetric2}
                \\
                \phi_{c,i}^\star &= \beta \,\phi_{c,i-1}^\star + (1-2\beta)\phi_{c,i}^\star + \beta \,\phi_{c,i+1}^\star + \beta (\boldsymbol{e}_n)_i\,, \,\,\, i\neq1\,.
                \label{Eq:ProofMultiChannelAsymmetric1}                         
        \end{align}
        From~\eqref{Eq:ProofMultiChannelAsymmetric2}, and since~$\gamma > 0$, we have~$\phi_{c,1}^\star = \phi^{*}_{c+1,1}$, for all~$c$ modulo~$C$. This makes the second summation term in~\eqref{eq:problemMulichannel} equal to zero, that is, $\boldsymbol{w_{c+1}}^T \boldsymbol{\phi}^\star_{c+1} = \boldsymbol{w_c}^T \boldsymbol{\phi}_c^\star$, for all~$c$. From~\eqref{Eq:ProofMultiChannelAsymmetric1}, and since~$\beta > 0$, we have
        \begin{align}
                \phi_{c,i}^\star &= \frac{\phi_{c,i-1}^\star + \phi_{c,i+1}^\star}{2}\,,\,\,\, i = 2,3,\ldots,n-1
                \label{Eq:ProofMultiChannelAsymmetricExtra}                             
                \\
                \phi_{c,n}^\star &= \frac{\phi_{c,n-1}^\star + \phi_{c,1}^\star + 1}{2}\,.
                \label{Eq:ProofMultiChannelAsymmetricExtra2}                            
        \end{align}             
        These equations are equivalent to $\phi_{c,i+1}^\star-\phi_{c,i}^\star = 1/n$, for~$i =1,\ldots,n-1$, and $\phi^\star_{c,1}+1-\phi^\star_{c,n} = 1/n$, and this makes the first term of the objective of~\eqref{eq:problemMulichannel} equal to zero. To see why the above equivalence holds, note that~\eqref{Eq:ProofMultiChannelAsymmetricExtra}-\eqref{Eq:ProofMultiChannelAsymmetricExtra2} imposes that all~$n-1$ phases~$\phi_{c,2}^\star$, $\phi_{c,3}^\star$, $\ldots$, $\phi_{c,n}^\star$ be placed in the interval~$[\phi_{c,1}^\star, \phi_{c,1}^\star+1]$. Furthermore, each phase has to equal the average of the previous phase with the next phase, where the phase previous to~$\phi_{c,2}^\star$ is $\phi_{c,1}^\star$ and the phase next to~$\phi_{c,n}$ is $\phi_{c,1}^\star+1$. The only possibility is all phases, including the extreme points, being equispaced.
        
        We now prove that~$\{\boldsymbol{\phi}^{(k)}\}$ converges. Writing~\eqref{Eq:MultiChannelAsymmetricIteration} in a more compact form,
        \begin{equation}\label{Eq:ProofMultiChannelAsymmetric3}
                \boldsymbol{\phi}^{(k)} = \boldsymbol{M}\boldsymbol{\phi}^{(k-1)} + \boldsymbol{b}\,.
        \end{equation} 
        It is known that the sequence~$\{\boldsymbol{\phi}^{(k)}\}$ produced by~\eqref{Eq:ProofMultiChannelAsymmetric3} converges to~$(\boldsymbol{I}-\boldsymbol{M})^{-1}\boldsymbol{b}$ whenever the spectral radius of~$\boldsymbol{M}$, denoted as~$\rho(\boldsymbol{M})$, is strictly smaller than~$1$~\cite[\S1.2]{Kelley95-IterativeMethodsLinearNonlinearEquations}. In our case, however, $1$ is an eigenvalue of~$\boldsymbol{M}$, so~$\rho(\boldsymbol{M}) \geq 1$. By computing all the eigenvalues of~$\boldsymbol{M}$, we will see that actually $\rho(\boldsymbol{M}) = 1$. Before proceeding, note that the vector of ones, $\boldsymbol{1}_{nC}$, is a right eigenvector of~$\boldsymbol{M}$ associated to the eigenvalue~$1$, and
        $
                \boldsymbol{u} := (\boldsymbol{e}_1, \boldsymbol{e}_1, \ldots, \boldsymbol{e}_1) \in (\mathbb{R}^{n})^C
        $
        is a left eigenvector of~$\boldsymbol{M}$ also associated to the eigenvalue~$1$.\footnote{If the Perron-Frobenuis theory~\cite{horn2012matrix,Meyer00-MatrixAnalysisAndAppliedLinearAlgebra} were applicable, we would conclude that~$\rho(\boldsymbol{M}) = 1$, and that the eigenvalue~$1$ would have algebraic multiplicity~$1$. This would enable us to skip the computation of all eigenvalues of~$\boldsymbol{M}$ and jump to the next paragraph. However, the Perron-Frobenuis theory is not applicable, since~$\boldsymbol{M}$, although being positive, is not irreducible.} To compute the eigenvalues of~$\boldsymbol{M}$, first   decompose~$\boldsymbol{Q_1}$ as
        $$
                \boldsymbol{Q_1}
                =
                \begin{bmatrix}
                        1-\gamma & \boldsymbol{0}_{n-1}^T \\
                        \boldsymbol{r} & \boldsymbol{T}
                \end{bmatrix}\,,
        $$
        where~$\boldsymbol{r} = (\beta, 0, \ldots, 0, \beta) \in \mathbb{R}^{n-1}$, and
        $$
                \boldsymbol{T}
                =
                \begin{bmatrix}
                        1-2\beta & \beta & 0 & \cdots & 0 & 0 \\
                        \beta & 1 -2\beta & \beta & \cdots & 0 & 0 \\
                        \vdots & \vdots & & \ddots & \cdots & \cdots \\
                        0 & 0 & & \cdots & \beta & 1-2\beta
                \end{bmatrix}\,.
        $$
        There is exists a permutation matrix~$\boldsymbol{P}$ such that
        $$
                \boldsymbol{P}^T\boldsymbol{M}\boldsymbol{P}
                =
                \left[
                \begin{array}{cccccccc}
                        \boldsymbol{T} &   &        &   & \boldsymbol{r}&   &        &   \\
                          & \boldsymbol{T} &        &   &  & \boldsymbol{r} &        &   \\
                          &   & \ddots &   &  &   & \ddots &   \\
                          &   &        & \boldsymbol{T} &  &   &        & \boldsymbol{r} \\
                          \multicolumn{4}{c}{\boldsymbol{0}_{C\times (n-1)C}} & \multicolumn{4}{c}{\boldsymbol{R}}
                \end{array}
                \right]\,,
        $$
        where 
        $$
        \boldsymbol{R}
        =
        \begin{bmatrix}
                1-\gamma & \gamma & 0 & \cdots & 0 \\
                0 & 1-\gamma & \gamma & \cdots & 0 \\
                \vdots & & \ddots & & \vdots  \\
                \gamma & 0 & 0 & \cdots & 1-\gamma   
        \end{bmatrix}\in \mathbb{R}^{C\times C}\,.
        $$
        Such permutation matrix corresponds to a reordering of the nodes such that $\boldsymbol{\phi}$ is mapped onto 
        \begin{multline*}
                (\phi_{1,2}, \phi_{1,3}, \ldots, \phi_{1,n}, \phi_{2,2}, \phi_{2,3}, \ldots, \phi_{2,n}, \ldots, \\\phi_{1,1}, \phi_{2,1}, \ldots, \phi_{C,1})\,,
        \end{multline*}
        that is, the first nodes of each channel are in the end of the vector in the new coordinate system. The matrices $\boldsymbol{M}$ and $\boldsymbol{P}^T\boldsymbol{M}\boldsymbol{P}$ have the same eigenvalues. The upper triangular structure of~$\boldsymbol{P}^T\boldsymbol{M}\boldsymbol{P}$ reveals that its eigenvalues are the roots of
        $
                \text{det}(\boldsymbol{T}-\lambda \boldsymbol{I})^{C}\text{det}(\boldsymbol{R}-\lambda \boldsymbol{I}) = 0\,,
        $
        where~$\boldsymbol{I}$ is an identity matrix with appropriate dimensions. In other words, the eigenvalues of~$\boldsymbol{M}$ are the union of the eigenvalues of~$\boldsymbol{T}$, each with multiplicity~$C$, with the eigenvalues of~$\boldsymbol{R}$, each with multiplicity~$1$. Since~$\boldsymbol{T}$ is tridiagonal Toeplitz, its eigenvalues are $\lambda_j(\boldsymbol{T}) = 1-2\beta + 2\beta \cos(\frac{\pi}{n}j)$, for $j=1,\ldots,n-1$~\cite[p.514]{Meyer00-MatrixAnalysisAndAppliedLinearAlgebra}. The matrix~$\boldsymbol{R}$, on the other hand, is a circulant matrix and hence its eigenvalues are the Fourier transform of the vector that generates the matrix. In this case, they are~$\lambda_j(\boldsymbol{R})=1-\gamma + \gamma \exp(\frac{2\pi i}{C}j)$, for~$j = 1, \ldots, C$, where~$i:= \sqrt{-1}$. Since~$0<\gamma<1$, $\boldsymbol{R}$ has one eigenvalue equal to~$1$ (multiplicity~$1$) and the remaining ones have magnitude smaller than~$1$. As~$0<\beta<1/2$, all eigenvalues of~$\boldsymbol{T}$ have magnitude smaller than~$1$. We conclude that~$\rho(\boldsymbol{M}) = 1$, and that its algebraic (and geometric) multiplicity is~$1$. 
        
        Define~$\overline{\boldsymbol{M}} := \boldsymbol{M} - \boldsymbol{1}_{nC}\boldsymbol{u}^T$. Then, $\rho(\overline{\boldsymbol{M}}) <1$~\cite[Lemma 8.2.7]{horn2012matrix}, and~\eqref{Eq:ProofMultiChannelAsymmetric3} can be written as 
        \begin{equation}\label{Eq:ProofMultiChannelAsymmetric4}
                \boldsymbol{\phi}^{(k)} = \overline{\boldsymbol{M}}\boldsymbol{\phi}^{(k-1)} + \boldsymbol{1}_{nC}\boldsymbol{u}^T \boldsymbol{\phi}^{(k-1)} + \boldsymbol{b}\,.
        \end{equation}
        Since~$\boldsymbol{u}^T \boldsymbol{M} = \boldsymbol{u}$ and ~$\boldsymbol{u}^T\boldsymbol{b} = 0$, \eqref{Eq:ProofMultiChannelAsymmetric3} tells us that $\boldsymbol{u}^T\boldsymbol{\phi}^{(k)} = \boldsymbol{u}^T\boldsymbol{M} \boldsymbol{\phi}^{(k-1)} + \boldsymbol{u}^Tb = \boldsymbol{u}^T \boldsymbol{\phi}^{(k-1)}$. In particular, 
        $$
        \boldsymbol{u}^T\boldsymbol{\phi}^{(k)} = \boldsymbol{u}^T\boldsymbol{\phi}^{(k-1)} = \cdots = \boldsymbol{u}^T\boldsymbol{\phi}^{(1)} = \boldsymbol{u}^T\boldsymbol{\phi}^{(0)}\,.
        $$
        Defining~$\overline{\boldsymbol{b}} = \boldsymbol{b} + \boldsymbol{1}_{nC}\boldsymbol{u}^T \boldsymbol{\phi}^{(0)}$, \eqref{Eq:ProofMultiChannelAsymmetric4} can then be written as
        \begin{equation}\label{Eq:ProofMultiChannelAsymmetric5}
                \boldsymbol{\phi}^{(k)} = \overline{\boldsymbol{M}}\boldsymbol{\phi}^{(k-1)} + \overline{\boldsymbol{b}}\,,
        \end{equation} 
        where~$\rho(\overline{\boldsymbol{M}})<1$. Thus, according to~\cite[\S1.2]{Kelley95-IterativeMethodsLinearNonlinearEquations}, the sequence $\{\boldsymbol{\phi}^{(k)}\}$ produced by~\eqref{Eq:ProofMultiChannelAsymmetric5}, and thus by~\eqref{Eq:MultiChannelAsymmetricIteration}, converges to~$(\boldsymbol{I}-\overline{\boldsymbol{M}})^{-1}\overline{\boldsymbol{b}} = (\boldsymbol{I}-\overline{\boldsymbol{M}})^{-1}\boldsymbol{b} + (\boldsymbol{I}-\overline{\boldsymbol{M}})^{-1}\boldsymbol{1}_{nC}\boldsymbol{u}^T \boldsymbol{\phi}^{(0)}$, which is well-defined and unique (note that~$\boldsymbol{I}-\overline{\boldsymbol{M}}$ is invertible because~$\rho(\overline{\boldsymbol{M}})<1$). This shows that the sequence~$\{\boldsymbol{\phi}^{(k)}\}$ converges.
\end{IEEEproof}


\bibliographystyle{IEEEtran}
\bibliography{refs}

\begin{thebibliography}{10}
\providecommand{\url}[1]{#1}
\csname url@samestyle\endcsname
\providecommand{\newblock}{\relax}
\providecommand{\bibinfo}[2]{#2}
\providecommand{\BIBentrySTDinterwordspacing}{\spaceskip=0pt\relax}
\providecommand{\BIBentryALTinterwordstretchfactor}{4}
\providecommand{\BIBentryALTinterwordspacing}{\spaceskip=\fontdimen2\font plus
\BIBentryALTinterwordstretchfactor\fontdimen3\font minus
  \fontdimen4\font\relax}
\providecommand{\BIBforeignlanguage}[2]{{%
\expandafter\ifx\csname l@#1\endcsname\relax
\typeout{** WARNING: IEEEtran.bst: No hyphenation pattern has been}%
\typeout{** loaded for the language `#1'. Using the pattern for}%
\typeout{** the default language instead.}%
\else
\language=\csname l@#1\endcsname
\fi
#2}}
\providecommand{\BIBdecl}{\relax}
\BIBdecl

\bibitem{deligiannis2015decentralized}
N.~Deligiannis, J.~F. Mota, G.~Smart, and Y.~Andreopoulos, ``Decentralized
  multichannel medium access control: Viewing desynchronization as a convex
  optimization method,'' in \emph{Proc. 14th International Conference on
  Information Processing in Sensor Networks (IPSN'15)}.\hskip 1em plus 0.5em
  minus 0.4em\relax ACM, 2015, pp. 13--24.

\bibitem{degesys2008towards}
J.~Degesys and R.~Nagpal, ``Towards desynchronization of multi-hop
  topologies,'' in \emph{Proc. IEEE Int. Conf. Self-Adaptive and
  Self-Organizing Syst. (SASO)}, 2008, pp. 129--138.

\bibitem{watteyne2012openwsn}
T.~Watteyne, X.~Vilajosana, B.~Kerkez, F.~Chraim, K.~Weekly, Q.~Wang,
  S.~Glaser, and K.~Pister, ``Openwsn: a standards-based low-power wireless
  development environment,'' \emph{Transactions on Emerging Telecommunications
  Technologies}, vol.~23, no.~5, pp. 480--493, 2012.

\bibitem{vilajosana2013realistic}
X.~Vilajosana, Q.~Wang, F.~Chraim, T.~Watteyne, T.~Chang, and K.~Pister, ``A
  realistic energy consumption model for {TSCH} networks,'' \emph{IEEE Sensors
  J.}, 2013.

\bibitem{tinka2010decentralized}
A.~Tinka, T.~Watteyne, and K.~Pister, ``A decentralized scheduling algorithm
  for time synchronized channel hopping,'' in \emph{Ad Hoc Netw.}, 2010, pp.
  201--216.

\bibitem{pagliari2011scalable}
R.~Pagliari and A.~Scaglione, ``Scalable network synchronization with
  pulse-coupled oscillators,'' \emph{IEEE Trans. Mobile Comput.}, vol.~10,
  no.~3, pp. 392--405, 2011.

\bibitem{BuranapanichkitAndreopoulos}
D.~Buranapanichkit and Y.~Andreopoulos, ``Distributed time-frequency division
  multiple access protocol for wireless sensor networks,'' \emph{IEEE Wirel.
  Comm. Lett.}, vol.~1, no.~5, pp. 440Ð--443, Oct. 2012.

\bibitem{IEEE802.15.4e-2012}
{IEEE 802.15.4e-2012}, ``{IEEE Standard for Local and Metropolitan Area
  Networks. Part 15.4: Low-Rate Wireless Personal Area Networks (LRWPANs)
  Amendment 1: MAC Sublayer},'' \emph{{IEEE Std.}}, Apr. 2012.

\bibitem{simeOne2008distributed}
O.~Simeone, U.~Spagnolini, Y.~Bar-Ness, and S.~H. Strogatz, ``Distributed
  synchronization in wireless networks,'' \emph{IEEE Signal Process. Mag.},
  vol.~25, no.~5, pp. 81--97, Sep. 2008.

\bibitem{patel2007desync}
A.~Patel, J.~Degesys, and R.~Nagpal, ``Desynchronization: The theory of
  self-organizing algorithms for round-robin scheduling,'' \emph{Proc. IEEE
  Int. Conf. Self-Adaptive and Self-Organizing Syst. (SASO)}, july 2007.

\bibitem{motskin2009lightweight}
A.~Motskin, T.~Roughgarden, P.~Skraba, and L.~Guibas, ``Lightweight coloring
  and desynchronization for networks,'' in \emph{IEEE INFOCOM'09}, 2009, pp.
  2383--2391.

\bibitem{lien2012anchored}
C.-M. Lien, S.-H. Chang, C.-S. Chang, and D.-S. Lee, ``Anchored
  desynchronization,'' in \emph{Proc. IEEE INFOCOM'12}, 2012, pp. 2966--2970.

\bibitem{leidenfrost2009firefly}
R.~Leidenfrost and W.~Elmenreich, ``Firefly clock synchronization in an
  802.15.4 wireless network,'' \emph{EURASIP J. Embed. Syst.}, 2009.

\bibitem{klinglmaye2012selforganizing}
J.~Klinglmayr and C.~Bettstetter, ``Self-organizing synchronization with
  inhibitory-couples oscillaotrs: convergence and robustness,'' \emph{ACM
  Trans. on Autonomous and Adaptive Systems}, vol.~7, no.~3, Sep. 2012.

\bibitem{hong2005scalable}
Y.-W. Hong and A.~Scaglione, ``A scalable synchronization protocol for large
  scale sensor networks and its applications,'' \emph{IEEE J. Sel. Areas
  Commun.}, vol.~23, no.~5, pp. 1085--1099, 2005.

\bibitem{choochaisriArtificialForceField}
S.~Choochaisri, K.~Apicharttrisorn, K.~Korprasertthaworn, P.~Taechalertpaisarn,
  and C.~Intanagonwiwat, ``Desynchronization with an artificial force field for
  wireless networks,'' \emph{ACM SIGCOMM Computer Communication Review},
  vol.~42, no.~2, pp. 7--15, 2012.

\bibitem{bojic2012self}
I.~Bojic, V.~Podobnik, I.~Ljubi, G.~Jezic, and M.~Kusek, ``A self-optimizing
  mobile network: Auto-tuning the network with firefly-synchronized agents,''
  \emph{Information Sciences}, vol. 182, no.~1, pp. 77--92, 2012.

\bibitem{mirollo1990synchronization}
R.~E. Mirollo and S.~H. Strogatz, ``Synchronization of pulse-coupled biological
  oscillators,'' \emph{SIAM Journal on Applied Mathematics}, vol.~50, no.~6,
  pp. 1645--1662, 1990.

\bibitem{scaglione2010bioinspired}
R.~Pagliari, Y.-W.~P. Hong, and A.~Scaglione, ``Bio-inspired algorithms for
  decentralized round-robin and proportional fair scheduling,'' \emph{IEEE J.
  on Select. Areas in Commun.}, vol.~28, no.~4, pp. 564--575, May 2010.

\bibitem{wang2012energy}
Y.~Wang, F.~Nunez, and F.~J. Doyle, ``Energy-efficient pulse-coupled
  synchronization strategy design for wireless sensor networks through reduced
  idle listening,'' \emph{IEEE Trans. Signal Process.}, vol.~60, no.~10, pp.
  5293--5306, 2012.

\bibitem{wang2012optimal}
Y.~Wang and F.~J. Doyle, ``Optimal phase response functions for fast
  pulse-coupled synchronization in wireless sensor networks,'' \emph{IEEE
  Trans. Signal Process.}, vol.~60, no.~10, pp. 5583--5588, 2012.

\bibitem{papachristodoulou2005synchronization}
A.~Papachristodoulou and A.~Jadbabaie, ``Synchronization in oscillator
  networks: Switching topologies and non-homogeneous delays,'' in \emph{IEEE
  Conf. Dec. Control (CDC'05)}, 2005, pp. 5692--5697.

\bibitem{olfati2007consensus}
R.~Olfati-Saber, J.~A. Fax, and R.~M. Murray, ``Consensus and cooperation in
  networked multi-agent systems,'' \emph{Proceedings of the IEEE}, vol.~95,
  no.~1, pp. 215--233, 2007.

\bibitem{simeone2007distributed}
O.~Simeone and U.~Spagnolini, ``Distributed time synchronization in wireless
  sensor networks with coupled discrete-time oscillators,'' \emph{EURASIP J.
  Wireless Commun. Netw.}, vol. 2007.

\bibitem{buranapanichkit2015convergence}
D.~Buranapanichkit, N.~Deligiannis, and Y.~Andreopoulos, ``Convergence of
  desynchronization primitives in wireless sensor networks: A stochastic
  modeling approach,'' \emph{IEEE Trans. Signal Process.}, vol.~63, no.~1, pp.
  221Ð--233, 2015.

\bibitem{Erseghe11-FastConsensusByADMM}
T.~{Erseghe}, D.~{Zennaro}, E.~{Dall'Anese}, and L.~{Vangelista}, ``Fast
  consensus by the alternating direction multipliers method,'' \emph{IEEE
  Trans. Signal Process.}, vol.~59, no.~11, pp. 5523--5537, 2011.

\bibitem{Mota12-ConsensusOnColoredNetworks-CDC}
J.~F. Mota, J.~M. Xavier, P.~M. Aguiar, and M.~Puschel, ``{D-ADMM}: A
  communication-efficient distributed algorithm for separable optimization,''
  \emph{IEEE Trans. Signal Process.}, vol.~61, no.~10, pp. 2718--2723, 2013.

\bibitem{nesterov1983method}
Y.~Nesterov, ``A method of solving a convex programming problem with
  convergence rate {$O(1/k^2)$},'' \emph{Soviet Mathematics Doklady}, vol.~27,
  no.~2, pp. 372--376, 1983.

\bibitem{Nesterov04-IntroductoryLecturesConvexOptimization}
Y.~{Nesterov}, \emph{Introductory Lectures on Convex Optimization: A Basic
  Course}.\hskip 1em plus 0.5em minus 0.4em\relax Kluwer Academic Publishers,
  2004.

\bibitem{refdegdesync}
J.~Degesys, I.~Rose, A.~Patel, and R.~Nagpal, ``Desync: self-organizing
  desynchronization and tdma on wireless sensor networks,'' in \emph{Int. Conf.
  on Information Processing in Sensor Networks (IPSN)}, 2007, pp. 11--20.

\bibitem{DeGroot74-ReachingConsensus}
M.~{DeGroot}, ``Reaching a consensus,'' \emph{J. American Statistical
  Association}, vol.~69, no. 345, pp. 118--121, 1974.

\bibitem{Boyd04-FastLinearIterationsforDistributedAveraging}
L.~{Xiao} and S.~{Boyd}, ``Fast linear iterations for distributed averaging,''
  \emph{Systems and Control Letters}, vol.~53, pp. 65--78, 2004.

\bibitem{rabbat2004distributed}
M.~Rabbat and R.~Nowak, ``Distributed optimization in sensor networks,'' in
  \emph{Int. Conf. Information Processing in Sensor Networks (IPSNÕ04)}.\hskip
  1em plus 0.5em minus 0.4em\relax ACM, 2004, pp. 20--27.

\bibitem{Vandenberghe11-Gradient-lecs}
L.~{Vandenberghe}, ``Gradient method,'' Spring 2008-09, lecture Notes,
  Optimization Methods for Large-Scale Systems (EE-236C), UCLA.

\bibitem{PerformanceanalysisGTS}
G.~Lu, B.~Krishnamachari, and C.~Raghavendra, ``Performance evaluation of the
  {IEEE} 802.15. 4 {MAC} for low-rate low-power wireless networks,'' in
  \emph{IEEE Internat. Conf. on Perf., Comput., and Comm.}, 2004, pp. 701--706.

\bibitem{wang20136tsch}
Q.~Wang, X.~Vilajosana, and T.~Watteyne, ``{6TSCH} operation sublayer (6top),''
  \emph{Internet-Draft, IETF Std., Rev. draft-wang- 6tisch-6top-sublayer-00},
  Apr. 2014.

\bibitem{deligiannis2014progressively}
N.~Deligiannis, F.~Verbist, J.~Slowack, R.~v.~d. Walle, P.~Schelkens, and
  A.~Munteanu, ``Progressively refined wyner-ziv video coding for visual
  sensors,'' \emph{ACM Trans. Sensor Netw.}, vol.~10, no.~2, p.~21, 2014.

\bibitem{boano2011jamlab}
C.~A. Boano, T.~Voigt, C.~Noda, K.~Romer, and M.~Z{\'u}{\~n}iga, ``Jamlab:
  Augmenting sensornet testbeds with realistic and controlled interference
  generation,'' in \emph{Int. Conf. on Information Processing in Sensor
  Networks (IPSN)}, 2011, pp. 175--186.

\bibitem{bertsekas1999nonlinear}
D.~P. Bertsekas, ``Nonlinear programming,'' 1999.

\bibitem{Spielman-SpectralGrapphTheory-lecs}
D.~{Spielman}, ``The {Laplacian},'' 2009, lecture notes, Spectral Graph Theory,
  Yale.

\bibitem{Kelley95-IterativeMethodsLinearNonlinearEquations}
C.~T. {Kelley}, \emph{Iterative Methods for Linear and Nonlinear
  Equations}.\hskip 1em plus 0.5em minus 0.4em\relax SIAM, Philadelphia, 1995.

\bibitem{horn2012matrix}
R.~A. Horn and C.~R. Johnson, \emph{Matrix analysis}.\hskip 1em plus 0.5em
  minus 0.4em\relax Cambridge university press, 2012.

\bibitem{Meyer00-MatrixAnalysisAndAppliedLinearAlgebra}
C.~M. {Meyer}, \emph{Matrix Analysis and Applied Linear Algebra}.\hskip 1em
  plus 0.5em minus 0.4em\relax SIAM, Philadelphia, 2000.

\end{thebibliography}

\end{document}